\documentclass[smallextended]{svjour3}
\usepackage{graphicx,bm, mathptmx}
\usepackage{subfigure}

\newcommand{\be}{\begin{equation}}
\newcommand{\ee}{\end{equation}}
\newcommand{\bea}{\begin{eqnarray}}
\newcommand{\eea}{\end{eqnarray}}
\newcommand{\beaa}{\begin{eqnarray*}}
\newcommand{\eeaa}{\end{eqnarray*}}

\newcommand{\nn}{\nonumber}

\def\rf#1{(\ref{#1})}

\def\prjnab{ \widetilde{\nabla}}
\def\confnab{\overline{\nabla}}
\def\prjconfnab{ \widetilde{\overline{\nabla}}}
\usepackage{amsmath,amssymb}
\def \curl {\mbox{curl}\,}
\begin{document}

\title{Conformal Transformations in Cosmology of Modified Gravity: the Covariant Approach Perspective}

\author{ Sante Carloni \and Emilio Elizalde \and Sergei Odintsov}

\institute{ Sante Carloni \at  Institut d'Estudis Espacials de Catalunya
(IEEC), Campus UAB, Facultat Ci\`{e}ncies, Torre C5-Par-2a pl, E-08193 Bellaterra
(Barcelona) Spain,\\ \email{carloni@ieec.uab.es}\\
Emilio Elizalde \at Consejo Superior de Investigaciones Cient\'{\i}ficas
(ICE/CSIC) \, and Institut d'Estudis Espacials de Catalunya
(IEEC),Campus UAB, Facultat Ci\`{e}ncies, Torre C5-Par-2a pl, E-08193 Bellaterra
(Barcelona) Spain,\\ \email{elizalde@ieec.fcr.es}\\
Sergei Odintsov \at ICREA, Barcelona, Spain \, and \,
ICE (CSIC-IEEC),
Campus UAB, Facultat Ci\`{e}ncies, Torre C5-Par-2a pl, E-08193 Bellaterra
(Barcelona) Spain,\\ \email{odintsov@ieec.fcr.es}}

\maketitle
\begin{abstract}
The 1+3 covariant approach and the covariant gauge-invariant approach to perturbations are used to analyze in depth conformal transformations in cosmology. Such techniques allow us  to obtain insights on the physical meaning of these transformations when applied to non-standard gravity. The results obtained lead to a number of general conclusions on the change of some key quantities describing any two conformally related cosmological models.   For example, even if some of the geometrical properties of the cosmology are preserved (homogeneous and isotropic Universes are mapped into homogeneous and isotropic universes), it can happen that decelerating cosmologies can be mapped into accelerated ones. From the point of view of the cosmological perturbations it is shown how these fluctuation transform.  We find that first-order vector and tensor perturbations equations are left unchanged in their structure by the conformal transformation, but this cannot be said of the scalar perturbations, which present differences in their evolutionary features.  The results obtained are then explicitly interpreted and verified with the help of some clarifying examples based on $f(R)$-gravity cosmologies.
\end{abstract}

%\tolerance=5000

%%%%%%%%%%%%%%%%%%%%%%%%%%%%%%%%%%%%%%%%%%%%%%%%%%%%%%%%%%
\section{Introduction}
%%%%%%%%%%%%%%%%%%%%%%%%%%%%%%%%%%%%%%%%%%%%%%%%%%%%%%%%%%

Conformal or Weyl transformations have played for long time an important role in many fields, from geography
(e.g.  Mercator projection) to electromagnetism  (see e.g. the work of Bateman \cite{Bateman}) to quantum theories \cite{Kastrup:2008jn}. In relativity and cosmology conformal transformations are also widely used. For example they can be exploited to introduce the so-called conformally flat spacetimes, which are among the simplest possible nontrivial spacetimes compatible with General Relativity (GR) \cite{J.L.Synge:1960zz}, or to construct Penrose diagrams  \cite{Hawking:1973uf}, which are one of the most important techniques for the study of black hole physics.

These transformations are also particularly important when one deals with non-standard theories of gravity. Such theories have recently been focus of much investigation because  they are thought to offer a possible explanation for the problem of Dark Energy (for recent reviews see \cite{Nojiri:2006ri,Capozziello:2007ec,Sotiriou:2008rp} ). One of the main difficulties in dealing with these theories, however, is that the non-linearity of their structure makes it really hard to deal with them. Hence much effort has been put into developing new methods to analyze those models. Conformal transformations are particularly convenient in this respect. In fact,  it is well known that with their use one is able to map non-standard theories of gravity into general relativity (GR) plus a scalar field ($\phi$) which is minimally coupled to the geometry \cite{Wands:1993uu,Maeda:1988ab}. For many years the meaning and the physical interpretation of these maps have been debated and nowadays one cannot say that a complete agreement has been reached in the community on this issue (see e.g. \cite{Magnano:1993bd,Flanagan:2004bz,Cotsakis:2004zg,Sotiriou:2007zu,Capozziello:2006dj,Dabrowski:2008kx} for some recent papers). In what follow we will not attempt to enter in this debate, but we will simply try to give a different perspective on the problem based on a new technique: the 1+3 covariant approach.

The 1+3 covariant approach has been developed in recent years thanks to pioneering work of Ehlers \cite{Ehlers1+3}, subsequently developed by Ellis (see for example \cite{EllisCovariant}). This formalism allows a treatment of any cosmological spacetime in a way that is, at the same time, mathematically rigorous and physically clear, and that can be easily adapted to non-standard theories of gravitation. The 1+3 covariant approach has been successful not only in the direct analysis of complicated cosmological models but also in other applications. One of them, the development of the dynamical system approach \cite{ellisbook}, has revealed itself very useful in shedding light on the dynamics of Bianchi Universes and the cosmology of scalar tensor and higher-order gravity (see \cite{ellisbook,Carloni:2004kp,Carloni:2007eu} for details).

It is a matter of fact that an important part of our knowledge on the cosmic evolution comes from the analysis of the perturbations of homogeneous and isotropic cosmological models. There are many different methods to describe the evolution of these perturbations, the most popular approach being Bardeen's one \cite{bardeen}. Classical reviews of this approach can be found, e.g., in \cite{Mukhanov:1990me,Kodama:1985bj}. In this paper, however, we will use the so-called Covariant Gauge Invariant  (CoGI) approach \cite{BDE,EB,EBH,DBE,BED}. This technique, based on the 1+3 approach, preserves the most appealing properties of the 1+3 formalism and allows the description of the first order perturbation of any spacetime. It has revealed itself crucial in the development of a consistent theory of perturbations in $f(R)$-gravity, as well as in other extensions of General Relativity \cite{SantePert,StructForm}. Differently from the other formulations \cite{Pogosian:2007sw,Tsujikawa:2007tg,Li:2008ai,HuSawicki,Bertschinger:2008zb,HuSawicki2,Nesseris:2008mq,delaCruzDombriz:2008cp,Motohashi:2009qn,DeFelice:2009ak},  the CoGI approach offers not only the simplest way to describe the evolution of the perturbations, but, as we will see, it also allows an easy comparison between the perturbations in the different frames.

The purpose of this paper is to look at conformal transformations  from the point of view of  the 1+3 approach and  the CoGI approach. These formalism will allow to see the conformal transformation under an entire different view, which hopefully might be useful for further investigations of this topic. We will discover that conformal transformation can be associated to a passage from an inertial observer to an accelerated one plus a redefinition of fields, and that a real change in the physics can occur only at the stage of the field redefinition. Also we will give general formulas that relate the main quantities in cosmology in the different frames including the transformation of the perturbations and this will shed light on the difference in their behavior in the different frames. In order to show that in a concrete example, we will apply this transformation to some models of $f(R)$-gravity. 

The paper is structured as follows. In Section II we give a brief review of the common procedure to perform a conformal transformation and we apply it to $f(R)$-gravity, with a specific focus on the distinction between change in the geometry and field redefinition, respectively. In Section III we introduce briefly the basics of the 1+3 covariant approach. In Section IV this formalism is applied to the conformal transformations and to the transformation of the covariant equations. Section V is dedicated to the corresponding transformation in the CoGI formalism.  In Section VI  the behavior of the scalar perturbations in the two frames is compared in detail for two simple models based on $f(R)$-gravity. Finally, Section VII is devoted to conclusions.

We now fix the notations. Unless otherwise specified, natural units
($\hbar=c=k_{B}=8\pi G=1$)
will be used throughout the paper, Latin indices running from 0 to 3.
The symbol $\nabla$ represents the usual covariant derivative and
$\partial$ corresponds to partial differentiation. We use the
$-,+,+,+$ signature and the Riemann tensor is defined by
\begin{equation}
R^{a}{}_{bcd}=W^a{}_{bd,c}-W^a{}_{bc,d}+ W^e{}_{bd}W^a{}_{ce}-
W^f{}_{bc}W^a{}_{df}\;,
\end{equation}
where the $W^a{}_{bd}$ are the Christoffel symbols (symmetric in
the lower indices), defined by
\begin{equation}
W^a_{bd}=\frac{1}{2}g^{ae}
\left(g_{be,d}+g_{ed,b}-g_{bd,e}\right)\;.
\end{equation}
The Ricci tensor is obtained by contracting the {\em first} and the
{\em third} indices
\begin{equation}\label{Ricci}
R_{ab}=g^{cd}R_{cadb}\;.
\end{equation}
Symmetrization and antisymmetrization  over the indices of a tensor are defined as
\begin{equation}
T_{(a b)}= \frac{1}{2}\left(T_{a b}+T_{b a}\right)\;,\qquad \qquad T_{[a b]}= \frac{1}{2}\left(T_{a b}-T_{b a}\right)\,.
\end{equation}
Finally, the Hilbert--Einstein action in the presence of matter is
given by
\begin{equation}
{\cal A}=\int d x^{4} \sqrt{-g}\left[R+ 2\mathcal{L}_{m}\right]\;.
\end{equation}

%%%%%%%%%%%%%%%%%%%%%%%%%%%%%%%%%%%%%%%%%%%%%%%%%%%%%%%%%
\section{Conformal transformations in relativity}
%%%%%%%%%%%%%%%%%%%%%%%%%%%%%%%%%%%%%%%%%%%%%%%%%%%%%%%%%
In this section we will give a basic introduction on the usual treatment of conformal transformations in Riemannian geometry, following mainly \cite{Wald:1984rg,Faraoni:1998qx,Magnano:1993bd}.
 Given a  spacetime $(M, g_{ab})$ with dimension $m$ ($m \geq 2 $), a conformal (or Weyl) transformation is defined as a transformation of the metric tensor given by
\be\label{CT}
g_{ab} \rightarrow \bar{g}_{ab}=\Upsilon g_{ab} \; ,
\ee
where  $\Upsilon =\Upsilon(x)$ is a regular, strictly positive function of the spacetime coordinates. This type of transformation does not affect the index structure and, as a consequence, preserves the angles between geometrical objects. In addition, since $d\bar{s}=\Upsilon ds$, it is clear that this transformation leaves the  causal structure unchanged, i.e.  null geodesics and  light cones are preserved\footnote{For a rigorous proof of this fact see \cite{Wald:1984rg}.}.

It is easy to derive the transformation rule for the determinant of the metric tensor $g$
\begin{equation}
\bar{g} \equiv \mbox{det}\left(
\bar{g}_{ab}\right) = \Upsilon^{m}g\;,
\end{equation}
where we have used the fact that $\bar{g}^{ab}=
\Upsilon^{-1}g^{ab}$. This allows us to derive the transformation laws for the Christoffel symbols, the Riemann and Ricci tensors, the Ricci scalar, and the Weyl tensor \cite{J.L.Synge:1960zz,birrell,Wald:1984rg,Faraoni:1998qx}:
\begin{equation} \label{TCGamma}
\bar{\Gamma}^{a}_{bc}=
\Gamma^{a}_{bc}+ \digamma^{a}_{bc} =\Gamma^{a}_{bc}+\frac{1}{2 \Upsilon}\left(
2\delta^{a}_{(b} \nabla_{c)}\Upsilon -g_{bc}\nabla^{a} \Upsilon \right)\; ,
\end{equation}
\begin{eqnarray} \label{TCRiem}
& & \bar{ R}_{abc}^{~~~d}=R_{abc}^{~~~d}-2 \nabla_{[a} \digamma^{d}_{b]c}+2 \digamma^{e}_{c[a} \digamma^{d}_{b]e}
%{R_{abc}}^{d}+\delta^{d}_{[a} \nabla_{b]}\nabla_{c} ( \ln \Upsilon )-g^{ds} g_{c [ a}\nabla_{b ]}\nabla_{s}( \ln \Upsilon ) +\frac{1}{2} \nabla_{[ a} ( \ln \Upsilon ) \delta^{d}_{b ]}\nabla_{c}( \ln \Upsilon ) \nonumber \\& & -\frac{1}{2}\nabla_{[ a}( \ln \Upsilon )g_{b ]c} g^{d s} \nabla_{s}( \ln \Upsilon )-\frac{1}{2}g_{c [ a} \delta^{d}_{b ]} g^{s r}\nabla_{s} ( \ln \Upsilon ) \nabla_{r} ( \ln \Upsilon )
\;,
\end{eqnarray}
\begin{eqnarray} \label{TCRicci}
& \bar{R}_{ab }=&R_{ab }-\frac{(m-2)}{2} \nabla_{a}\nabla_{b }( \ln \Upsilon )
-\frac{1}{2} g_{ab } \Box( \ln \Upsilon )+\frac{(m-2)}{4}   \nabla_{a} ( \ln \Upsilon ) \nabla_{b}( \ln \Upsilon )\nonumber \\
&& -\frac{(m-2)}{4}  g_{ab }\, g^{ r s}
\nabla_{r}( \ln \Upsilon )  \nabla_{s}( \ln \Upsilon ) \; ,
\end{eqnarray}
\begin{eqnarray} \label{TCRicciSca}
\bar{R} \equiv \bar{g}^{ab} \bar{R}_{ab }=\frac{1}{\Upsilon} \left[ R- \left( m-1 \right) \Box \left( \ln \Upsilon \right) -
\frac{\left( m-1 \right) \left( m-2 \right)}{4}\frac{g^{ab} \nabla_{a} \Upsilon \nabla_{b}\Upsilon}{\Upsilon^{2}}
\right] \;,
\end{eqnarray}
\be
\bar{C}{_{abc}}^{d}={C_{abc}}^{d}\quad \mbox{(only with this index configuration)}\; ,
\ee
where $\Box= g^{r s } \nabla_{r} \nabla_{s}$.

An equation is said to be {\it conformally invariant} if there exists a number, $s$, such that if $\varPsi$ is a solution of this equation then $\bar{\varPsi}=\Upsilon^s\varPsi$ is a solution of its conformally transformed one \cite{Wald:1984rg}.

In relativity and cosmology, with the term conformal transformations one usually means two different transformations: (i) the transformation \rf{CT} on the metric and (ii) a rearrangement of the basic quantities in the theory. For example,  a Friedmann--Lema\^{\i}tre-Robertson--Walker (FLRW) metric with flat spatial sections
 \be   \label{EdS1}
ds^2= -dt^2+a^2 ( t )\left( dx^2+dy^2+dz^2 \right) \; ,
\ee
can be written as
\be   \label{EdS2}
ds^2=a^2 ( t ) \left( -\frac{dt^2}{a^2(t)}+dx^2+dy^2+dz^2 \right) \; ,
\ee
and, defining  the conformal time $d\eta$ as $dt/a(t)$, one can write
 \be   \label{EdS3}
ds^2=a^2 ( \eta ) \left( -d\eta^2+dx^2+dy^2+dz^2 \right) \; .
\ee
We will see that exactly the same happens when we apply the conformal transformation to a theory of gravity: the transformation of the metric will be associated to a re-parametrization of the fields in the action. In fact the latter is a crucial point in  understanding this application of the conformal transformations, because it is only in the last step that real changes in the model, which are unrelated to the geometrical conformal transformation \rf{CT}, take place. In the following  we will call {\it Jordan frame} (JF) the initial set of metric and fields present in the theory and {\it Einstein frame} (EF) the set of metric and fields obtained after this conformal transformation \footnote{This definition is by no means commonly used in literature, where different aspect of the theories (like the presence and position of non minimal couplings in the action) or other arguments have been used to define the conformal frames. The reason behind our choice is that, in principle one can choose as starting point a theory with any feature, the structure depending on the observations and the data coming from experiments, and this choice is in fact irrelevant to the understanding of the features of the conformal transformations.}. 

%%%%%%%%%%%%%%%%%%%%%%%%%%%%%%%%%%%%%%%%%%%%%%%%%%%%%%%%%%%
\subsection{Conformal Transformations and Higher Order Gravity}\label{CF-HOG}
%%%%%%%%%%%%%%%%%%%%%%%%%%%%%%%%%%%%%%%%%%%%%%%%%%%
Let us now see in detail what happens to the theory of gravity when we apply
the conformal transformation (\ref{CT}) to the metric tensor. As said above, we will focus on the so-called $f(R)$-gravity, which is among the most studied models of modified gravity.  Before starting, it is
worth to point out that one can apply a conformal transformation at different levels, namely: (i) at the action level or (ii) at the level of the field equations. Of course, since these transformations are, in principle, different from each other, one has to prove their equivalence, but, as we will see, this can be verified directly.

%%%%%%%%%%%%%%%%%%%%%%%%%%%%%%%%%%%%%%%%%%%%%%%%%%%%%%%%%%%
\subsubsection{Conformal transformations at the action level}
%%%%%%%%%%%%%%%%%%%%%%%%%%%%%%%%%%%%%%%%%%%%%%%%%%%
Let us start at the action level. A general Lagrangian for $f(R)$-gravity is given by
\begin{equation}\label{lagr f(R)}
L=\sqrt{-g}\left[ f(R)+2{\cal L}_{m}(\gimel,g^{ab})\right]\;,
\end{equation}
$\gimel$ being a generic matter field\footnote{It is worth to specify that here we are performing  a conformal transformation of the metric only, leaving the matter fields (as well as any other additional field) untransformed. This assumption is widely used in the literature and, for the sake of simplicity, we will adopt it here too.}. This action reduces to the Hilbert-Einstein one for $f(R)=R$. It is interesting to note that, performing the transformations (\ref{TCRiem}-\ref{TCRicciSca}) directly, with the action in this form, would result in a very complicated expression. Instead, using the tools of analytical mechanics \cite{Magnano:1993bd,Magnano:1988tw,Magnano:1990qu}, one manages to reduce the action to the so-called {\it Helmoltz form}, by means of defining a set of new fields associated with the higher-order momenta in the Lagrangian. A more pragmatic way to do that is to write  the action as \cite{Faraoni:1998qx,Nojiri:2003ft,Wands:1993uu}
\begin{equation}\label{HeLagfR}
 L=\sqrt{-g}\left[ A(R-B)+ f(B)+2{\cal L}_{m}(\gimel,\bar{g}^{ab})\right]\;,
\end{equation}
where $A$ and $B$ are two auxiliary fields related to the canonical momenta of \rf{lagr f(R)}. As a consequence one can prove that, on the one side, variation upon the field $A$ implies $B=R$, which means that this action is equivalent to \rf{lagr f(R)}, and on the other, that variation upon $B$ implies, instead, that $A$ and $B$ are related by $A=f'(B)$ \cite{Nojiri:2003ft}.

The Lagrangian \rf{HeLagfR} is effectively a scalar-tensor Lagrangian and can, consequently, be conformally transformed in the same way as in these theories. Specifically, from \rf{CT} one obtains
\begin{eqnarray}\label{LAG-f(R) Conformal Frame}
 &&\nn\bar{L}=\sqrt{-\bar{g}} \;\Upsilon^{-2}\left\{ \Upsilon F(B) \left[ \bar{R}+ 3 \overline{\Box} \left( \ln \Upsilon \right) -
\frac{3}{2}\frac{\bar{g}^{ab} \confnab_{a} \Upsilon \confnab_{b}\Upsilon}{\Upsilon^{2}}
\right]\right.\\ &&\left.\quad-F(B)B+{f}(B)+2{\cal L}_{m}(\gimel,\bar{g}^{ab})\right\}\;,
\end{eqnarray}
being $F=df/dB$ and where we have used the transformation laws\footnote{Note that the conformal transformation performed here is limited to the metric tensor. Since the field $B$ defined in \rf{HeLagfR} is a  generic field, one can set $\bar{B}=B$.} \rf{TCGamma}-\rf{TCRicciSca}.

At this point one proceeds to the reparametrization of the fields in the action. Setting\footnote{Note the absolute value in this definition. It is required in order to preserve the causal structure of the metric ($\Upsilon$ has to be positive) and implies that the conformal transformation can be only done on sections of the total history of the systems in which $F(B)$ has a constant sign.}
\begin{equation}\label{UpsilonDEF}
 \Upsilon= \left|F(B)\right|\;,
\end{equation}
one obtains
\begin{equation}\label{PreConf-LAG-f(R)}
 \bar{L}=\sqrt{-\bar{g}} \left[ \bar{R}%+ 3 \overline{\Box} \left( \ln \Upsilon \right)
-\frac{3}{2}\frac{\bar{g}^{ab} \confnab_{a} \Upsilon \confnab_{b}\Upsilon}{\Upsilon^2}
-2U(\Upsilon)+2\Upsilon^{-2}{\cal L}_{m}(\gimel,\bar{g}^{ab})\right]\;,
\end{equation}
where
\begin{equation}\label{PotTCAction}
U(\Upsilon)=\left.\frac{BF(B)- \bar{f}(B)}{2F(B)^2}\right|_{B= F^{-1}(\gamma\Upsilon/2)}\;,\end{equation}
$\gamma=\bar{|F|}/F$, and we have used the Gauss theorem to eliminate the term $3 \overline{\Box} \left( \ln \Upsilon \right)$.

The Einstein frame can be then achieved by redefining  $\Upsilon$ as $\Upsilon= e^{\lambda \phi}$ and considering $\phi$ a new (scalar) field in the theory. This step allows us to eliminate the non-minimal coupling in the kinetic term of \rf{PreConf-LAG-f(R)} and to obtain:
\begin{equation}\label{LAG-f(R)-EF}
 \bar{L}=\sqrt{-\bar{g}} \left[ \bar{R} - \bar{g}^{ab} \confnab_{a} \phi \confnab_{b}\phi
-2W(\phi)+2e^{-2 \sqrt{2/3}\phi }{\cal L}_{m}(\gimel,\bar{g}^{ab})\right]\;,
\end{equation}
once $\lambda$ is chosen to be $\sqrt{2/3}$. In this form, this Lagrangian looks like  a standard GR + minimally-coupled-scalar-field theory, and in  vacuum this would be indeed the case. However,  the {\it matter} Lagrangian in \rf{LAG-f(R)-EF}  appears to be coupled with the scalar field. This is the reason why in the literature it is often mentioned that in the Einstein frame matter is non-minimally coupled to the scalar field $\phi$. The presence of such coupling is understandable if we bear in mind that $\phi$ is in fact a part of the gravitational interaction in the JF and, as such, it is bound to have a direct coupling with standard matter. To wit, a non-minimal coupling between gravity and matter is already present  implicitly in \rf{lagr f(R)}, so that one can imagine that the conformal transformation separates fourth-order gravity into a tensorial part, which is minimally coupled with matter, and a scalar part, which carries the non-minimal coupling. In this sense the non-minimal coupling is an expression of the universality of the gravitational interaction in the Jordan frame. 

%%%%%%%%%%%%%%%%%%%%%%%%%%%%%%%%%%%%%%%%%%%%%%%%%%%%%%%%%%%
\subsubsection{Conformal transformations at the field equation level}
%%%%%%%%%%%%%%%%%%%%%%%%%%%%%%%%%%%%%%%%%%%%%%%%%%%

Let us now look at the transformation \rf{CT} from the point of view of the gravitational field equations. Upon variation, the Lagrangian \rf{lagr f(R)} gives rise to field equations that can be recast as \cite{Revnostra}:
\begin{eqnarray}\label{eq:field-f(R)}
 G_{ab}=\frac{T_{ab}^{m}(\gimel,\bar{g}^{ab})}{F}+T^{R}_{ab}=T^{tot}_{ab}\:,
 \end{eqnarray}
 where the term
 \begin{eqnarray}\label{Ttot}
 T^{R}_{ab}=\frac{1}{ F}g_{ab} \left(f-R F\right) +
\frac{\nabla_c\nabla_d F}{F}(g_{a}^{c}g_{b}^{d}-g_{ab}g^{cd})\;,
\end{eqnarray}
can be considered to represent an effective fluid associated with the non-Einstein contributions to the gravitational interaction, the term $\displaystyle{T^{m}_{ab}=\frac{2}{\sqrt{-g}}\frac{\delta(L_{m})}{\delta g_{ab}}}$ represents the
stress-energy tensor of standard matter, $\displaystyle{F(R)=\frac{df}{dR}}$, and we have dropped the $R$-dependence of $f$ and $F$. Also these equations reduce to the standard Einstein field equations when $f(R)=R$.

Using \rf{TCRicci} and \rf{TCRicciSca} directly on the L.H.S. of \rf{eq:field-f(R)}, we obtain the Einstein-tensor transformation law
\begin{eqnarray}\label{GabConfEqn}
&&\nn{G}_{ab}=\bar{G}_{ab} + \frac{1}{2} \confnab_a \ln(\Upsilon)\confnab_b \ln(\Upsilon)+ \frac{1}{4}\bar{g}_{ab}\bar{g}^{cd}  \confnab_c \ln(\Upsilon)\confnab_d \ln(\Upsilon)\\ &&\qquad+ \confnab_a  \confnab_b \ln(\Upsilon)-\bar{g}_{ab}\overline{\Box}\ln(\Upsilon)\,.
\end{eqnarray}
At this point supposing  $F=\bar{F}$ \cite{Magnano:1988tw,Magnano:1990qu,Magnano:1993bd}, one can transform the energy momentum  tensor $T^{tot}_{ab}$, obtaining\footnote{Note that in the operation just mentioned we have left unchanged the Ricci scalar $R$. This happens because one considers $R=R(F)$ and $F$ is left unchanged by the transformation. This is analogous to the introduction of the field $B$ in the Lagrangian derivation.}
\begin{eqnarray}
&&\nn  \bar{T}^{tot}_{ab}= \frac{1}{F}\bar{T}_{ab}^{m}(\gimel,\bar{g}^{ab}) -\bar{g}_{ab}\;\bar{U}(F)+ \confnab_a  \confnab_bF-g_{ab}\overline{\Box}F\\ &&\qquad- \confnab_cF  \confnab_d\Upsilon\left(\bar{g}^{c}_{(a}\bar{g}^{d}_{b)}+\frac{1}{2}\bar{g}^{ab}\bar{g}_{cd}\right)\:,
\end{eqnarray}
where
\begin{equation}\label{PotTCEqn1}
\bar{U}(F)=\left.-\frac{1}{2}\left(\frac{f-RF}{ F^2}\right)\right|_{R=R(F)}\;.
\end{equation}

In order to pass to the Einstein frame, we first need to set  $\Upsilon=F$ and then to introduce a scalar field $\phi$  such that  $\Upsilon=e^{\lambda\phi}$. In this way
 \begin{equation}
  {G}_{ab}=\bar{G}_{ab} + \frac{\lambda^{2}}{2} \confnab_a \phi\confnab_b \phi+ \frac{\lambda^{2}}{4}\bar{g}_{ab}\bar{g}^{cd}  \confnab_c \phi\confnab_d \phi+ \lambda\confnab_a  \confnab_b \phi-\lambda\bar{g}_{ab}\Box\phi\,,
\end{equation}
 \begin{eqnarray}
 &&\nn \bar{T}^{tot}_{ab}= e^{-\lambda\phi}\bar{T}_{ab}^{m}(\gimel,\bar{g}^{ab}) -\bar{g}_{ab}W(\phi) +2\lambda^{2}\confnab_a\phi  \confnab_b\phi\\&&\qquad-\frac{\lambda^{2}}{2} \confnab_c\phi  \confnab^c\phi+\lambda \confnab_a  \confnab_b\phi- \lambda \bar{g}_{ab}\overline{\Box}\phi\:,
\end{eqnarray}
where the scalar field potential is defined as
\begin{equation}\label{PotTCEqn2}
W(\phi)=\left.\bar{U}\right|_{F=e^{\lambda\phi}}\;,
\end{equation}
which is equivalent to \rf{PotTCAction}.

At this point, setting $\lambda=\sqrt{2/3}$ one obtains
\begin{equation}\label{FieldEqEinst}
\bar{G}_{ab}= e^{-\lambda\phi}\bar{T}_{ab}^{m}(\gimel,\bar{g}^{ab})+\confnab_a\phi  \confnab_b\phi-\frac{1}{2}\bar{g}_{ab} \confnab_c\phi  \confnab^c\phi-g_{ab}W(\phi)\:.
\end{equation}
Equations \rf{FieldEqEinst} describe  Einstein gravity plus a scalar field minimally coupled with gravity and non minimally coupled with standard matter. This theory coincides with the one directly derived upon variation of \rf{LAG-f(R)-EF}.  Such result shows that the conformal transformation \rf{CT}, with
\begin{equation}\label{phiDEF}
F=\Upsilon=\exp\left(\sqrt{2/3}\phi\right)\;,
\end{equation}
leads, both at the action and field equation levels, to the ``same" theory in the Einstein frame.

Comparing \rf{FieldEqEinst} and \rf{eq:field-f(R)} it is clear that the possibility to perform a conformal transformation has many advantages if one deals with a matter-less system. However, if matter is added the conformal transformation does not necessarily lead to an easier model. This is due mainly to the non-minimal coupling between standard matter and $\phi$ appearing in \rf{FieldEqEinst}  and \rf{LAG-f(R)-EF}, which induces additional terms in the Bianchi identities. For example, the Klein Gordon equation for $\phi$ reads
 \begin{equation}
\left[\overline{\Box} \phi-V'(\phi))\right]\prjconfnab{}_{c}\phi =\frac{1}{\sqrt{6}} \exp\left(-\sqrt{2/3}\phi\right) \bar{g}^{ab}\bar{T}^{m}_{ab}\prjconfnab{}_{c}\phi\;,
\end{equation}
and the energy-momentum conservation is given by
\begin{equation}
 \prjconfnab{}^b \bar{T}^{m}_{b a}= \sqrt{\frac{2}{3}}\bar{T}^{m}_{b a} \prjconfnab{}^{b}\phi -\frac{1}{\sqrt{6}}  \bar{T}^{m}\prjconfnab{}_{a}\phi \,.
\end{equation}
The above equations tell us that only a form of matter-energy for which the trace $\bar{T}$ is null renders the above equation conformally invariant.

In spite of its usefulness, the conformal transformations of theories of gravity bring a serious problem: the possibility of changing
the type and number of the fields in a theory by a simple change in the metric tensor implies that there is no reason, a priori, to choose a specific representation of the action among all possible ones. In other words, recognizing the freedom associated to the conformal mapping means, in fact, loosing the physics of the theory in an infinite set of representations. This fact, in itself, would not be a problem if those representations  would describe the same physics but, as we will see later, this does not appear to be the case: they describe very different Universes. As a consequence 
we are left with the choice of either establish the existence of the particular ``physical frame" (i.e. the specific field parametrization, that reflects the {\it actual} physical fields), or to prove that somehow all the frames are equivalent. There is a wide literature on this issue and we will not enter into the details of the debate referring the reader to some of the many papers and reviews on the topic (see e.g. \cite{Faraoni:1998qx,Magnano:1993bd,Flanagan:2004bz,Sotiriou:2007zu}). The purpose of this paper is to offer, using the covariant approaches, a new perspective on conformal transformation that might contribute to the clarification of this issue.

%%%%%%%%%%%%%%%%%%%%%%%%%%%%%%%%%%%%%%%%%%%%%%%%%%%%%%%%%%
\section{The  1+3 covariant approach to cosmology }\label{CoVApp}
%%%%%%%%%%%%%%%%%%%%%%%%%%%%%%%%%%%%%%%%%%%%%%%%%%%%%%%%%%
In this section we will present a brief introduction to the  covariant approach to cosmology. We will use this approach to  understand better the physics behind the conformal transformations and for the construction of a theory of cosmological perturbations in the two frames.

Given a  space-time associated to a cosmological model  one can single out
 a family of preferred worldlines representing a certain class of observers (for example the ones comoving with standard matter). If we suppose that it is possible to define a unique 4-velocity vector field  $u^a$  associated  to these worldlines, then we can split the metric tensor as
\be\label{gdecom1+3}
g_{ab} = h_{ab} - u_a\,u_b\:,
\ee
i.e. the spacetime is foliated in hypersurfaces with metric $h_{ab}$ orthogonal to the vector field $u_{a}$. In this way any affine parameter on the worldlines associated to  $u_a$ can be chosen to represent ``time" and the tensor $h_{ab}$  ($h^a{}_c\,h^c{}_b = h^a{}_b \ ,~ h^a{}_a = 3 \ , ~h_{ab}\,u^b = 0$) determines the geometry of the instantaneous rest-spaces of the observers we have chosen.  Using $u_a$ and $h_{ab}$, one can then define the projected volume form $\eta_{abc}=u^d\eta_{abcd}$, the covariant time derivative ($\,\dot{}\,$) along the fundamental worldlines, and the  fully orthogonally projected covariant derivative $\prjnab$:
\be
\dot{X}^{ab}{}_{cd} = u^{e}\nabla_{e}X^{ab}{}_{cd} \ , \qquad
\prjnab_{e}X^{ab}{}_{cd} = h^a{}_f\,h^b{}_g\,h^p{}_c\,h^q{}_d\,
h^r{}_e\nabla_r\,X^{fg}{}_{pq} \ .
\ee
Performing a split of the first covariant derivative of $u_a$ into its irreducible parts, namely
\be
\label{eq:kin}
\nabla_{a}u_{b} =
-\,u_a\,a_b + {\frac{1}{3}}\,\Theta\,h_{ab} + \sigma_{ab}
+ \omega_{ab} \, ,
\ee
one can define the basic kinematical quantities of this formalism \cite{EllisCovariant}.
The trace $\Theta = \prjnab_au^a{}$ is the  rate
of volume expansion scalar of the worldlines of $u_a$ (which is proportional
to the standard Hubble parameter $H$: $H=3\Theta$); $\sigma_{ab} = \prjnab_{\langle a}u_{b\rangle}$
is the trace-free symmetric  rate of shear tensor
($\sigma_{ab} = \sigma_{(ab)}$, $\sigma_{ab}\,u^b = 0$,
$\sigma^a{}_{a}= 0$) describing the rate of distortion of the
observer flow; $\omega_{ab} = \prjnab_{[a}u_{b]}$ is the
skew-symmetric vorticity tensor ($\omega_{ab} =
\omega_{[ab]}$, $\omega_{ab}\,u^b = 0$)
describing the rotation of the observers relative to a non-rotating
(Fermi-propagated) frame, and $a_b=\dot{u}_b$ is the acceleration vector,
which describes the non-gravitational forces acting on the observers\footnote{For an introduction to relativistic fluid mechanics and more information on the meaning of these tensors we refer the reader to \cite{Ehlers1+3,Plebanski:2006sd}.}.

A general matter energy-momentum tensor $T_{ab}$ can  also be
decomposed locally using $u_a$ and $h_{ab}$. One has
\bea
T_{ab} = \mu\,u_a\,u_b + q_a\,u_b + u_a\,q_b + p\,h_{ab} + \pi_{ab}\;,
\label{eq:stress}
\eea
where $\mu = (T_{ab}u^{a}u^{b})$ is the relativistic energy
density relative to $u^a$, $q^{a} = -\,T_{bc}\,u^{b}\,h^{ca}$ ($q_a\,u^a = 0 $) is
the  relativistic momentum density, which is also the energy
flux relative to $u^a$, $p = {\frac{1}{3}}\,(T_{ab}h^{ab})$ is the
 isotropic pressure, and $\pi_{ab} = T_{cd}\,h^{c}{}_{\langle
a}\,h^{d}{}_{b\rangle}$  ($\pi^a{}_a = 0 \ , ~\pi_{ab} = \pi_{(ab)} $)
is the trace-free  anisotropic pressure.

The quantities presented above completely determine a cosmological model. Their evolution and constraint equations, also known as {\it 1+3 covariant equations}, are completely equivalent to the Einstein equations and characterize  the full evolution of the cosmology. They are shown in Appendix \ref{AppA}.  The advantages in using these variables is that they allow for a treatment of cosmology that is both mathematically rigorous and  physically meaningful  and they are particularly useful in the construction of the theory of perturbations.

%%%%%%%%%%%%%%%%%%%%%%%%%%%%%%%%%%%%%%%%%%%%%%%%%%%%%%%%%
\section{The 1+3 Conformal Transformation}
%%%%%%%%%%%%%%%%%%%%%%%%%%%%%%%%%%%%%%%%%%%%%%%%%%%%%%%%%%

As we have seen in the previous section, a conformal transformation in relativity and cosmology is, in fact, the combination of a geometric operation and a field redefinition. We will treat them separately.

%%%%%%%%%%%%%%%%%%%%%%%%%%%%%%%%%%%%%%%%%%%%%%%%%%%%%%%%%
\subsection{The geometric part of the conformal transformation}
%%%%%%%%%%%%%%%%%%%%%%%%%%%%%%%%%%%%%%%%%%%%%%%%%%%%%%%%%%
Let us  look at the geometric part of the conformal transformation in terms of the 1+3 covariant approach.  Starting from \rf{CT}  and using \rf{gdecom1+3}, we can write
\begin{equation}\label{TC_1+3}
{g}_{ab}\rightarrow \bar{g}_{ab}=\Upsilon g_{ab}%=\Upsilon h_{a b}- (\sqrt{\Upsilon}\, u_a)(\sqrt{\Upsilon}\,u_b)
\qquad \Rightarrow \qquad
\left\{
\begin{array}{l}
  {h}_{ab} \rightarrow  \bar{h}_{ab}= \Upsilon h_{a b} \;,  \\
  {u}_{a} \rightarrow  \bar{u}_{a}= \sqrt{\Upsilon}u_a\ .
\end{array}
\right.
\end{equation}
The equations above show how the fact that the conformal factor is positive translates in the fact that the conformal observer velocity is always well defined and has to have the same direction of the Jordan observer. In addition, $\Upsilon>0$ implies that the sign of the projector tensor $h_{ab}$ remains the same, preserving the pseudo-Riemannian character of the manifold. The relations above also imply that in terms of the 1+3 formalism a conformal transformation\footnote{In the following we will consider this transformation as a passive transformation for $g_{ab}$. The reason for doing so is physical. If the transformations were active, we would basically change spacetime and the comparison of two observers in two different spacetimes would be less physically consistent.}  can be associated to a change  from  the Jordan observer $O_J$, associated to $u_a$,  to a new one which we will call Conformal observer $O_C$, associated to $\bar{u}_a$\footnote{Consistently with the tradition in Relativity, here we call ``observer" a reference frame in a specific state of motion.}. In particular: ({\it i}) the conformal observer has a 4-velocity whose modulus depends on the spacetime coordinates (and as a consequence is accelerated), and  ({\it ii}) the spatial metric of this observer is modified by the conformal factor.  This  tells us that \rf{TC_1+3} basically consists in switching from an inertial observer to an observer whose  clock rate and rod length change continuously  in spacetime.  This can be seen clearly looking at the transformation of the derivative operators. For scalars, we have
\begin{gather}
X^{\scriptscriptstyle \dag} = \bar{g}^{a c} \bar{u}_{c} \overline{\nabla}_{\,a}X=\frac{1}{\Upsilon}{g}^{a c}\, \sqrt{\Upsilon}u_{a}\nabla^{\,c}X=\frac{1}{\sqrt{\Upsilon}}\ \dot{X}\,, \\
 \widetilde{\overline{\nabla}}_{e}X=  \bar{h}^{~r}_{e} \overline{\nabla}_{r}X= h^{~r}_{e}\nabla_{r}X= \widetilde{\nabla}_{e}X\,,
\end{gather}
for vectors
\begin{gather}\label{CovDerVecTrans}
\nn X^{\scriptscriptstyle \dag}_a =  \bar{g}^{c b} \bar{u}_{c}\overline{\nabla}_{\,b}X_a=\frac{1}{\Upsilon}{g}^{c b}\, \sqrt{\Upsilon}u_{c}(\nabla_{\,b}X_a-\digamma^{c}_{b a}X_c)\\ \qquad=\frac{1}{\sqrt{\Upsilon}}\left[\dot{X}_a -\frac{1}{\Upsilon}u^{b}X_{(b}\nabla_{a)} \Upsilon+\frac{1}{2\Upsilon}\, u_{a} X^{r}\nabla_r \Upsilon\right]\,,\\
 \nn\prjconfnab_{e}X_{ a} =  \bar{h}^{~ c}_{a}\;\bar{h}^{~r}_{e} \overline{\nabla}_{r}X_{c}= h^{~ f}_{a}\;h^{~r}_{e}(\nabla_{\,r}X_f-\digamma^{c}_{f r}X_c)\\ \qquad= \prjnab_{e}X_{ a}-\frac{1}{\Upsilon}X_{(e}\prjnab_{a)} \Upsilon+\frac{1}{2\Upsilon}h_{ea}X^{r}\nabla_{r} \Upsilon\,,
\end{gather}
and the ones for tensors follow accordingly\footnote{Note that when one changes the position of the index of $X$ the corrections $\digamma$ change their sign in the same way as for the Christoffel symbols.}.

Since the derivatives are changed, the basic quantities that one uses to describe the cosmology and the perturbations are also changed. The 1+3 kinematical quantities are transformed as follows
\begin{eqnarray}\label{TransfKin1+3}
&&\bar{\Theta}=\frac{1}{\sqrt{\Upsilon}} \left(\Theta+\frac{3}{2}\frac{\dot{\Upsilon}}{\Upsilon}\right)\,,\\
&&\bar{\sigma}_{ab}=\sqrt{\Upsilon} \sigma_{ab}\,,\\
&&\bar{\omega}_{ab}=\sqrt{\Upsilon} \omega_{ab}\,,\\
&&\bar{a}_{b}=a_{b}+ \frac{1}{2}\frac{\prjnab_b\Upsilon}{\Upsilon}\,,
\end{eqnarray}
and the electric and magnetic parts of the Weyl tensor are transformed in themselves (one should beware of the position of the indices):
\begin{eqnarray}
&&\overline{E}_{a b}={E}_{a b},\\
&& \overline{H}_{a b}={H}_{a b}.
\end{eqnarray}
The transformations above describe clearly the differences between $O_C$ and $O_J$. The conformal observer sees an expansion rate which is increased if the conformal factor grows in time and it might even observe the Universe undergoing cosmic acceleration when in the Jordan frame the expansion is decelerated\footnote{Similarly, when  the cosmology becomes singular in the Einstein frame the modified gravity description in the Jordan frame  shows qualitatively a different behavior (e.g. it might become a complex theory \cite{BrisceseBamba}). } \cite{Capozziello:2006dj}.    Instead, the vorticity and the shear are only changed by a multiplicative factor, so that under conformal transformation homogeneous, isotropic and irrotational universes do not loose their symmetries. Such effect can be traced back to the fact that the velocity of the conformal observer is always parallel to the one of the Jordan observer. Finally, the transformation of the acceleration vector shows that even if we start from Universes with zero acceleration (i.e. no additional forces other than gravity) the conformal observer perceives an acceleration which depends on the spatial dependence of the conformal factor.

The connection between conformal transformation and observers is also important when one looks at the thermodynamics. Since $O_C$ is moving with respect to $O_J$ with varying velocity, he/she will measure different thermodynamics. In fact, using the transformation \rf{TC_1+3} one can write the energy momentum tensor as
\begin{equation}
T_{ab}=\frac{\mu}{\Upsilon} \bar{u}_a \bar{u}_b+ \frac{p}{\Upsilon} \bar{h}_{a b}+\frac{2}{\sqrt{\Upsilon}} {q}_{(a}\bar{u}_{b)}+\pi_{ab}\;,
 \end{equation}
and the conformal observer will detect
\begin{subequations}\label{ThermCF}
\begin{flalign}
&\bar{\mu}=T_{ab} \bar{u}^a\bar{u}^b =\frac{\mu}{\Upsilon}\;,\\
& \bar{p}=\frac{1}{3}T_{ab}\bar{h}_{ab}= \frac{p}{\Upsilon}\;,\\
&\bar{q}_a= -\,T_{bc}\,\bar{u}^{b}\,\bar{h}^{ca}=\frac{{q}_{a}}{\sqrt{\Upsilon}}\;,\\
 &\bar{\pi}_{ab}=T_{cd}\,\bar{h}^{c}{}_{\langle
a}\,\bar{h}^{d}{}_{b\rangle}
 =\pi_{ab}\;.
\end{flalign}
\end{subequations}
Thus if we assume standard matter in the JF to be a perfect fluid in its rest frame, in the Einstein frame standard matter remains a perfect fluid in the conformal frame (i.e. $\bar{q}_a=0$, $\bar{\pi}_{a b}=0$). In fact, with these transformations, in general also the equation of state is preserved. However,  since $O_C$ will measure only the barred quantities, the spacetime variation of all the thermodynamical quantities is different. The thermodynamics is further modified by the transformations in the derivatives which lead to further changes in the usual conservation laws.  For example, in the homogeneous and isotropic cases one has
\begin{equation}
\bar{\mu}^\dag+\bar{\Theta}(\bar{\mu}+\bar{p})-\frac{1}{2} (\bar{\mu}+3\bar{p})  \frac{\Upsilon^\dag}{\Upsilon}=0\;.
\end{equation}
This is easy to understand  in terms of the properties of the observer described above. Since the rods of the conformal observer change in time and space, the mass energy contained in a box at rest with this observer will change,
and $O_C$ will measure a modification of the standard conservation laws.

At this point, using the transformation of the kinematics and the thermodynamics presented above, one can derive how the 1+3 equations transform under \rf{CT}. The detailed set of equations is given in Appendix \ref{AppA}. One can see that the conformal observer perceives many corrections to the standard cosmological equations. These equations show that for an accelerated observer, like $O_C$, the expressions appear deeply modified in their structure. It is interesting to note, {\it en passant},  that such observer, if unaware of its acceleration, would conclude that some exotic physics or change in the gravitational interaction is taking place on cosmological scale.

%%%%%%%%%%%%%%%%%%%%%%%%%%%%%%%%%%%%%%%%%%%%%%%%%%%%%%%%%
\subsection{The field-redefinition part of the conformal transformation}
%%%%%%%%%%%%%%%%%%%%%%%%%%%%%%%%%%%%%%%%%%%%%%%%%%%%%%%%%%

Let us now concentrate on the remaining part of the conformal transformation. As we have mentioned before, this consists basically in a field redefinition. In principle there is no standard prescription for the definition of a field in a theory, however the structure of the 1+3 equations and what we know about field theory (the correct form for a kinetic term, etc.) suggests the definition \rf{phiDEF} should be taken. This specific choice (or any other whatsoever) leads to a tremendous change in the model, which the 1+3 formalism helps appreciate in detail.

As we have seen, the conformal observer uses clocks and rods that change with the spacetime coordinates. This means that such observer will perceive, for example, an object moving with a constant velocity with respect to the Jordan observer as if it was accelerating. Associating the conformal factor to a scalar field basically amounts to considering such effects as a result of the presence of a new interaction, rather than a kinetic effect. In a way this resembles Einstein's lift  {\it Gedankenexperiment}: the (accelerated) conformal observer becomes an inertial observer, which we will call {\it Einstein observer} $O_E$, and a scalar field is introduced in the model which accounts for the additional kinematics.

What said above explains the fact that, if one performs a conformal transformation, even in pure General Relativity, one obtains Einstein's gravity plus a scalar field. It also allows to clarify the nature of $\phi$. Such field cannot be really considered a matter field, even if it behaves exactly like one: the best interpretation for $\phi$ is, in our view, to consider it as a kinematical effect promoted to interaction.

When one applies the conformal transformation to $f(R)$-gravity two additional operations are performed, namely the specification of the nature of the thermodynamical quantities and the connection of the scalar field with the $f(R)$ term.  Both steps require particular attention. Let us consider the first one. If we look at the 1+3 equations as perceived by $O_C$, we can see that these equations are different from the ones we would obtain for the theory \rf{LAG-f(R)-EF} even if we would write $\Upsilon$ in terms of $\phi$.  The reason is that these equations miss a critical ingredient, i.e. the specification of the structure of $\bar{\mu}$ which represent the {\it total} energy density as derived  from $T_{a b}^{tot}$ in \rf{Ttot}. In making this substitution one has to remember the presence of the non-minimal coupling between standard matter and the Ricci scalar and the fact that---differently to what happens with standard-matter thermodynamical variables---the effective variable associated to $T^{R}_{ab}$ contain derivative terms. This implies that the effective thermodynamic quantities associated to $T^{R}_{ab}$ do not follow strictly the transformations \rf{ThermCF}. In particular,
\begin{subequations}\label{ThermCFf(R)}
\begin{flalign}
&\bar{\mu}^R= \frac{1}{\Upsilon}\left(\mu^R+ F\left(1-\frac{1}{\Upsilon}\right) W(F)-\frac{3}{2}\frac{\dot{\Upsilon}}{\Upsilon}\frac{\dot{F}}{F} +\frac{1}{2}\frac{\prjnab_a F\prjnab{}^a \Upsilon}{F^2} \right)\;,\\
& \bar{p}^R= \frac{1}{\Upsilon}\left(p^R- F\left(1-\frac{1}{\Upsilon}\right) W(F)+\frac{1}{2}\frac{\dot{\Upsilon}}{\Upsilon}\frac{\dot{F}}{F}-\frac{5}{6}\frac{\prjnab_a F\prjnab{}^a \Upsilon}{F^2}  \right)\;,\\
&\bar{q}^R_a=\frac{1}{\sqrt{\Upsilon}}\left({q}^R_{a}-\frac{1}{2}\frac{\dot{\Upsilon}}{\Upsilon}\frac{\prjnab_a F}{F}-\frac{1}{2}\frac{\dot{F}}{F}\frac{\prjnab_a \Upsilon}{\Upsilon}\right)\;,\\
 &\bar{\pi}^R_{ab}=\pi_{ab}+\frac{\prjnab_{\langle a} F}{F}\frac{\prjnab{}_{b\rangle} \Upsilon}{\Upsilon} \;,
\end{flalign}
\end{subequations}
where
\begin{subequations}\label{Therm-f(R)}
\begin{flalign}
&\mu^{R}\,=\,\frac{1}{F}\left[\frac{1}{2}(R F-f)-\Theta
\dot{F}+\tilde{\nabla}^2{F}\right]\label{muR}\;,\\
&p^{R}\,=\,\frac{1}{F}\left[\frac{1}{2}(f-R
F)+\ddot{F}+\frac{2}{3}\Theta
\dot{F}-\frac{2}{3}\tilde{\nabla}^2{F}  -\,a_b\prjnab^b{F}\right]\label{pR}\;,\\
&q^{R}_a\,=\,-\frac{1}{F}\left[\prjnab_{a}\dot{F}-\frac{1}{3}\Theta
\prjnab_{a}F-\sigma_{a b}\prjnab^{b}F-\omega_{a b}\prjnab^{b}F\right]\;,\\
&\pi^{R}_{ab}\,=\,\frac{1}{F}\left[\prjnab_{\langle
a}\prjnab_{b\rangle}F-\sigma_{a b}\dot{F}\right]\,,\label{piR}\\
&W(F)=\frac{R(F) F-f(F)}{2F^{2}}\,.
\end{flalign}
\end{subequations}
Once the correct transformations are introduced, one can substitute
\begin{equation}\label{PhiDef-f(R)}
\left|F\right|=e^{\lambda\phi}\;,
\end{equation}
to obtain the equations one would derive from \rf{LAG-f(R)-EF}. Again, the 1+3 approach helps shedding light on the physical meaning of this important step. The relation between $F$
and $\phi$ modifies the cosmological equations in such a way that all the  higher-order terms are compensated, and one is just left with a linear theory of gravity and a scalar field minimally coupled to the geometry. In other words, one is thereby choosing a specific form of the conformal factor for which the kinematical terms compensate the non-Einstenian part of the equations. Thus, in practice, the Einstein observer moves in such a way to compensate the $f(R)$ correction. Such compensation is complete in vacuum, but constraints matter to move non geodetically (at least with the choice \rf{PhiDef-f(R)}) as a footprint of the transformation we performed. Also this fact bears clear similarities with Einstein's lift experiment: the only way in which the observer in the lift is able to infer the presence of an actual gravitational field is to study the geodesic deviation of matter.

%%%%%%%%%%%%%%%%%%%%%%%%%%%%%%%%%%%%%%%%%%%%%%%%%%%%%%%%%%
\section{Perturbations and conformal transformations}
%%%%%%%%%%%%%%%%%%%%%%%%%%%%%%%%%%%%%%%%%%%%%%%%%%%%%%%%%%
At this point we are ready to move our attention on how the conformal transformations affect the evolution of the cosmological perturbations. This will be done using the CoGI approach, which is based on the 1+3 equations mentioned in the previous section and listed in Appendix \ref{AppA}.

The CoGI approach presents one main difference (which is at the same time a point of strength) with respect to other perturbation theory approaches in that it relies directly on the structure of the perturbed Universe, rather than on the concepts of background quantities and perturbations. In normal cases, the structure of the perturbed spacetime is trivial because one can just consider a completely generic spacetime. However, when we want to compare the perturbations of two conformally related spacetimes, the structure of the perturbed Universe in the ``arrival" frame is not generic, but depends on the type of transformation chosen. In what follows we will assume that the conformal factor is a function of all the spacetime coordinates\footnote{ In principle one could choose a conformal factor which depends on spatial or temporal coordinates only, but this would induce problems in the connection between the conformal factor and the scalar field  made in Sect.~\ref{CF-HOG}. For example, performing a conformal transformation  with a conformal factor that depends, say, only on time, would result in the disappearance of all the projected derivatives of $\phi$ in the 1+3 equations  and would make impossible to characterize the perturbation of this field. If one would force the perturbations on these quantities, like one seems to be able to do in other perturbation formalisms, the perturbation of $\phi$ would represent a fluctuation of the conformal mapping, introducing something similar to gauge modes in  the theory.}.

The next step in the construction of the CoGI formalism is the definition of the  background. This is not done by assigning a metric, but rather by recognizing which 1+3 quantities are zero in the background and which are not. In what follows we will consider expanding ($\Theta\neq0$) homogeneous and isotropic ($\sigma_{ab}=0$, $\omega_{ab}=0$) backgrounds. In this setting we will characterize the perturbations in terms the of 1+3 quantities seen in Sect.~\ref{CoVApp} and their projected gradients. For example, the key quantities relevant to the evolution of scalar perturbations in GR are
\begin{equation}
 D_a =\frac{ S }{\mu}\prjnab_a\mu\;, \quad Z_a \equiv  S
\prjnab_a\Theta\;,
\quad C_a \equiv   S^3 \prjnab_a R_3\;, \label{eq:vardefDZC}
\end{equation}
which represent the comoving normalized spatial gradient of the energy density, the comoving spatial gradient of the expansion, and the comoving spatial gradient of the 3-Ricci scalar, respectively. These variables are related by a constraint coming form the spatial derivative of the Gauss equation \cite{SantePert,StructForm}.
Moreover it can be proven that these variables, as well as any other quantity which vanish in the background, are gauge-invariant \cite{bi:stewart}.

A quick look to the Einstein frame 1+3 equation in App.~\ref{AppA} shows clearly that the tensor and vector perturbation equations are left unchanged in their structure, but, as we will see, the same cannot be said of the scalar perturbations. In the following we will focus on this last type of perturbations only and, specifically, on spherically symmetric collapse, which is associated  to the cosmological density fluctuations. To extract this information from the variables \rf{eq:vardefDZC}, we use the local splitting 
\begin{equation}\label{localdecomposition}
S\prjnab_{a}X_b=X_{ab}= \frac{1}{3}h_{ab}X+\Sigma^{X}_{ab}+X_{[ab]}\;,
\end{equation}  where
\begin{equation}
\Sigma^{X}_{ab}=X_{(ab)}-\frac{1}{3}h_{ab}X\;,
\end{equation}
and we then single out  the scalar parts of   \rf{eq:vardefDZC}:
\begin{equation}\label{ScaVar}
\Delta^{m}=\frac{S^2}{\mu^{m}}\prjnab^2\mu^{m}\,,\qquad
Z=S^2\prjnab^2\Theta\,,\qquad C=S^{4}\prjnab^2\tilde{R}\;,
\end{equation}
which are gauge invariant, for the same reasons that $D_a, Z_a, C_a$ are. We will select (and deal) with variables of the type \rf{ScaVar} only.

In order to describe the scalar fluctuations in $f(R)$-gravity (as in any other gravitational theory) we will use the variables \rf{eq:vardefDZC}, plus other ones that will take into account the additional degrees of freedom of the theory and are defined specifically for the theory itself. The results of \cite{bi:stewart} will guarantee that these new quantities are indeed gauge invariant.

At this point it is relatively easy to construct the perturbation equations. Starting from the 1+3 equations one obtains a set of propagation and constraint equations for  the variables \rf{ScaVar}. Then, one chooses a background while recognizing which of these variables is zero in the background. These variables are then considered to be of order one. At this point the linearized equations can be obtained by dropping all the terms of order higher than one from the propagation and constraint equations.

Before analyzing in detail the transformation of the perturbation equations thus obtained, let us consider---as we already did for the kinematical quantities---what can we learn from the transformation of the perturbation variables \rf{ScaVar} upon \rf{CT}. We have
\begin{subequations}\label{GRPertVarTransf}
\begin{flalign}
&\overline{D}_a=\frac{\bar{S}}{\bar{\mu}}\prjconfnab_{a}\bar{\mu} =\sqrt{\Upsilon} \left(D_a- S\frac{\prjnab_{a}\Upsilon}{\Upsilon }\right)\label{GRPertVarTransf1} \,,\\
&\overline{Z}_a= {Z}_a-\frac{1}{2}S \Theta \frac{\prjnab_{a}\Upsilon}{\Upsilon }-\frac{9}{4}S \frac{\dot{\Upsilon}}{\Upsilon}\frac{\prjnab_{a}\Upsilon}{\Upsilon }+\frac{3}{2}S  \frac{\prjnab_{a}\dot{\Upsilon}}{\Upsilon }\;,\\
&\overline{C}_a=\sqrt{\Upsilon } \left[C_a-2 S^2 Z_a\frac{ \Upsilon ' }{\Upsilon}+S^3 \left(8 \Theta\frac{\dot{\Upsilon}}{\Upsilon }+9 \frac{\dot{\Upsilon}^2}{\Upsilon^2 }-2 \tilde{R}\right)\frac{\prjnab_a \Upsilon}{\Upsilon}\right.\nn
\\ &\left.- S^3\left(2 \Theta +\frac{3 \Upsilon '}{\Upsilon}\right) \frac{\prjnab_a\dot{\Upsilon}}{\Upsilon }\right].
\end{flalign}
\end{subequations}
This clearly shows that the matter fluctuations in the Einstein frame are a combination of the matter fluctuations in the Jordan frame with the fluctuations of the conformal factor. This can be understood, intuitively, if one thinks that the conformal observer measures the matter fluctuations with clocks and rods which are also perturbed.  It is useful to give the transformation for the scalar variables too, which read, to first order,
 \begin{subequations}\label{ScaPertVarTransf}
\begin{flalign}
&\overline{\Delta}=\Upsilon\left(\Delta- S^2\frac{\prjnab^2\Upsilon}{\Upsilon }\right)\label{ScaPertVarTransf1}\,,\\
&\overline{Z}= \sqrt{\Upsilon}\left({Z}-\frac{1}{2}S^{2} \Theta \frac{\prjnab^{2}\Upsilon}{\Upsilon }-\frac{9}{4}S^{2} \frac{\dot{\Upsilon}}{\Upsilon}\frac{\prjnab^2\Upsilon}{\Upsilon }+\frac{3}{2}S^{2}  \frac{\prjnab^2\dot{\Upsilon}}{\Upsilon }\right)  \;,\\
&\overline{C}=\sqrt{\Upsilon } \left[C-2 S^2 Z\frac{ \dot{\Upsilon } }{\Upsilon}+S^3 \left(8 \Theta\frac{\dot{\Upsilon}}{\Upsilon }+9 \frac{\dot{\Upsilon}^2}{\Upsilon^2 }-2 \tilde{R}\right)\frac{\prjnab^2 \Upsilon}{\Upsilon}\right.\nn
\\ &\left.- S^3\left(2 \Theta +\frac{3 \Upsilon '}{\Upsilon}\right) \frac{\prjnab^2\dot{\Upsilon}}{\Upsilon }\right].
\end{flalign}
\end{subequations}
The differences between the Jordan and the Einstein frames appear clearly: even if the JF matter fluctuations are close to zero for some reason, $O_E$ can still be able to observe matter fluctuations, or, depending on the choice of the conformal factor, in spite of the presence of  matter fluctuations in the Jordan frame  the conformal observer could possibly see no matter fluctuations at all! In addition,  because of the transformations above, it seems clear that $\dot{C}=0$ does {\it not} necessarily imply that $\dot{\bar{C}}=0$. This means that in the long wavelength limit the system of perturbation does not posses a conserved quantity, like it happens in GR.  Such feature will have an important impact on the difference in the perturbation behaviors in the two frames.

The equations above also show that, in general, the perturbation equations are not conformally invariant in the sense of \cite{Wald:1984rg}. For example, given the structure of \rf{GRPertVarTransf1} and \rf{ScaPertVarTransf1}, one can see that it would be difficult to prove that there exists a number $s$ such that $\overline{D}_a=\Upsilon^{s} D_a$.

%%%%%%%%%%%%%%%%%%%%%%%%%%%%%%%%%%%%%%%%%%%%%%%%%%%%%%%%%%
\subsection{Scalar perturbations of $f(R)$-gravity in the Jordan frame}
%%%%%%%%%%%%%%%%%%%%%%%%%%%%%%%%%%%%%%%%%%%%%%%%%%%%%%%%%%
Let us now derive explicitly the perturbation equations  for $f(R)$-gravity around an homogeneous and isotropic  background in the presence of a barotropic fluid with equation of state $p^m=\omega \mu^m$. The zeroth order equations are given by
\begin{subequations}
\begin{flalign}
&\Theta^2\,=\,3\frac{\mu^{m}}{F} + 3\mu^{R}-\frac{3R_3}{2}\;,\\
&\dot{\Theta}+{\textstyle\frac{1}{3}}\Theta^2
+{\textstyle\frac{1}{2 F}}({\mu}^{m} + 3{p}^{m})
        +{\textstyle\frac{1}{2}}({\mu}^{R} + 3{p}^{R})=0\;,\\
& \dot{\mu}^m\,+ \,\Theta\,(\mu^m+{p^m})=0\;,\\
& \dot{\mu}^R\,+ \,\Theta\,(\mu^R+{p^R})-\mu^m\frac{F'}{F^2}\dot{R}=0\;,
\end{flalign}
\end{subequations}
where $\mu^{R}$ and $p^{R}$ are given in \rf{muR} and \rf{pR}, $R_3$ is the 3-Ricci scalar and $R_3=6K/S^2$ with the spatial curvature index $K=0,\pm1$ and $S$  the scale factor.

Now, in order to model the additional degrees of freedom of this theories one can add to \rf{ScaVar} the following scalar quantities
\begin{equation}\label{PertVarJF}
{\cal R}=S^2\prjnab^2 R\,,\qquad\Re=S^2\prjnab^2 \dot{R}\;,
\end{equation}
where  ${\mathcal R}$  determine the fluctuations in the Ricci scalar  $R$  and  $\Re$ and the ones of its momentum $\dot{R}$\footnote{This choice of variables is by no means unique, but the ones we have chosen are definitely among the most convenient.}. Again, since these quantities vanish in the background, we can say that, as in the case of $\Delta, Z$ and $C$, they are gauge invariant. The set of variables $\Delta, Z, C, {\cal R},\Re$  completely characterizes the evolution of  the density perturbations in $f(R)$-gravity. Their evolution equations constitute a system of first order partial differential equations \cite{SantePert,StructForm}. In order to reduce it to a system of ordinary differential equations, one defines the eigenfunctions of the spatial Laplace-Beltrami operator:
\begin{equation}\label{CovHarmDef}
\prjnab^{2}Q = -\frac{\ell^{2}}{S^{2}}Q\,,
\end{equation}
where $\ell=2\pi S/\lambda$ is the wave number and $\dot{Q}=0$,  and expands every first order quantity in the above equations:
\begin{equation}\label{eq:developmentdelta}
X(t,\mathbf{x})=\sum X^{(\ell)}(t)\;Q^{(\ell)}(\mathbf{x})\;,
\end{equation}
where $\sum$ stands for both summation over discrete or integration over continuous indices. In this way, one obtains the equations describing the $\ell^{th}$ mode for scalar perturbations in $f(R)$ gravity. They are \cite{SantePert,StructForm}:
\begin{eqnarray}\label{ScaPertJF}
&&\dot{\Delta}_{m}^{(\ell)} =w\Theta \Delta_{m}^{(\ell)}-(1+w)Z^{(\ell)}\,,\label{eqDeltaHarm}\\
&&\dot{Z}^{(\ell)} = \left(\frac{\dot{R}
   F'}{F}-\frac{2 \Theta }{3}\right)Z^{(\ell)}+
    \left[\frac{ (w -1) (3 w +2)}{2 (w +1)} \frac{\mu^{m}}{ F} + \frac{2 w \Theta ^2
   +3 w (\mu^{R}+3  p^{R}) }{6 (w +1) }\right]   \Delta_{m}^{(\ell)}+\nonumber \\  &&+\frac{\Theta F'}{F}\Re^{(\ell)}+
   \left[\frac{1}{2}-\frac{ F'}{F} \frac{\ell^2}{S^2}-\frac{1}{2} \frac{f}{F}\frac{ F'}{F}- \frac{F'}{F} \frac{\mu^{m}}{
   F} + \dot{R} \Theta  \left(\frac{F'}{F}\right)^{2}+ \dot{R} \Theta \frac{ f^{(3)}}{ F}\right]\mathcal{R}^{(\ell)}\,,\\
&&\dot{{\cal R}}^{(\ell)}=\Re^{(\ell)}-\frac{w }{w +1}\dot{R}\;{\Delta}_{m}^{(\ell)}\,,\label{eqZHarm}\\
&&\dot{\Re}^{(\ell)}=- \left(\Theta + 2\dot{R} \frac{f^{(3)}}{F'}\right)\Re^{(\ell)}- \dot{R} Z^{(\ell)} -
  \left[\frac{ (3 w -1)}{3} \frac{\mu^{m}}{F'} + \frac{w}{3(w +1)} \ddot{R} \right]{\Delta}_{m}^{(\ell)}+\nonumber\\ &&+\left[\frac{\ell^{2}}{S^2}-\left(\frac{1}{3}\frac{F}{F'}+\frac{f^{(4)}}{F} \dot{R}^2+\Theta \dot{R} \frac{f^{(3)}}{F'}+\ddot{R} \frac{f^{(3)}}{F'}-\frac{R}{3}\right)\right]\mathcal{R}^{(\ell)}\label{eqRho2}\,.
\end{eqnarray}
These equations have been thoroughly studied in \cite{SantePert,StructForm}, we refer the reader to these papers for additional information on their properties.
%%%%%%%%%%%%%%%%%%%%%%%%%%%%%%%%%%%%%%%%%%%%%%%%%%%%%%%%%
\subsection{Scalar perturbations  of $f(R)$-gravity in the Einstein frame.}
%%%%%%%%%%%%%%%%%%%%%%%%%%%%%%%%%%%%%%%%%%%%%%%%%%%%%%%%%%
Let us consider now the Einstein frame\footnote{In the following we will reconstruct the perturbation equations from the 1+3 system given in Appendix \ref{AppA}. Of course one could have made the transformation directly from \rf{ScaPertJF} using the formulas given above. The result is, of course, the same.}. The Lagrangian and the general field equations are given by \rf{LAG-f(R)-EF} and \rf{FieldEqEinst} respectively. Considering the background choices in the Jordan frame and the transformations \rf{TransfKin1+3}, \rf{ThermCF} and \rf{ThermCFf(R)}, we obtain the associated background equations:
\begin{subequations}
\begin{flalign}
&\bar{\Theta}^2\,=\,3\bar{\mu}^{m}e^{\left(-\sqrt{2/3}\phi\right)} + 3\mu^{\phi}-\frac{3\bar{R}_3}{2}\;,\\
&\bar{\Theta}^\dag+{\textstyle\frac{1}{3}}\bar{\Theta}^2
+{\textstyle\frac{1}{2}}(\bar{\mu}^{m} + 3\bar{p}^{m})e^{\left(-\sqrt{2/3}\phi\right)}
+{\textstyle\frac{1}{2}}({\mu}^{\phi} + 3{p}^{\phi})=0\;,\\
& \bar{\mu}_m^\dag\,+ \,\bar{\Theta}\,(\bar{\mu}^m+\bar{p}^m)-\sqrt{\frac{2}{3}}\;\bar{\mu}^m\phi^\dag-\frac{1}{\sqrt{6}} (3\bar{p}^m-\bar{\mu}^{m})\phi^\dag=0\;,\\
& \overline{\Box} \phi-W'(\phi)=\frac{1}{\sqrt{6}} (3\bar{p}^m-\bar{\mu}^{m})\; e^{\left(-\sqrt{2/3}\phi\right)}\;,
\end{flalign}
\end{subequations}
where
\begin{subequations}
\begin{flalign}
&\mu^{\phi}\,=\frac{1}{2}(\phi^\dag)^2+\,\frac{1}{2}\prjconfnab\,{}^a\phi\prjconfnab_a \phi+W(\phi)\;,\\
&p^{\phi}\,=\frac{1}{2}(\phi^\dag)^2\,-\frac{1}{6}\prjconfnab\,{}^a\phi\prjconfnab_a\phi-W(\phi)\;.
\end{flalign}
\end{subequations}
In order to model the additional degrees of freedom, we need to  define two additional variables:
\begin{equation}\label{PertVarEF}
\bar{\Phi}=S^2\prjconfnab\,{}^2 \phi\,,\qquad\bar{\Psi}=S^2\prjconfnab\,{}^2 {\phi}^\dag\;,
\end{equation}
which represent, by construction \cite{bi:stewart}, the gauge invariant fluctuations of the scalar field and its momentum.  The relation between the variables in the Jordan and Einstein frames, at linear order, is given by
\begin{subequations}
\begin{flalign}
&\bar{\Delta}^m=F\left(\Delta^m-\frac{F'}{F}\mathcal{R}\right)\,,\\
&\bar{Z}=Z-\frac{1}{2}\Theta\frac{F'}{F}\mathcal{R}+\frac{3}{2}\frac{F'}{F}\Re \,,\\
&\bar{\Phi}=\sqrt{\frac{3}{2}}\frac{F'}{F}\mathcal{R}\,,\\
&\bar{\Psi}=\sqrt{\frac{3}{2 F}}\left[\left(\frac{F''}{F}-\frac{3}{2}\frac{(F')^2}{F^2}\right)\dot{R}\,\mathcal{R}+\frac{F'}{F}\Re\right]\,.
\end{flalign}
\end{subequations}
This allows us to connect the initial conditions in the two frames.
The fact that in the Einstein Frame we need the same number of variables as in the Jordan  one, shows that in the conformal transformation no information on the degrees of freedom is lost, as it is expected.

In performing the harmonic decomposition one needs to remember that the defining equation for the covariant harmonics \rf{CovHarmDef}  has to be transformed too, so that the $Q$s could be different. However, using \rf{CovDerVecTrans} one obtains
\begin{equation}
\Upsilon\prjconfnab{}^{2}\bar{Q} -\frac{1}{2}\frac{\prjconfnab{}^{a}\bar{Q}\prjconfnab_{a}\Upsilon}{\Upsilon}= -\Upsilon\frac{\ell^{2}}{\bar{S}^{2}}\bar{Q}\,,
\end{equation}
which shows that, at first order, $\bar{Q}=Q$. Using this, one is able to write the perturbation equations as follows
\begin{eqnarray}\label{ScaPertEF}
&&\nn\Delta^{\dag}_{(\ell)}= \left[w \Theta -\frac{w (3 w+1)\phi^{\dag}}{\sqrt{6} (w+1)}\right] \Delta_{(\ell)}-(w+1) Z_{(\ell)}+\frac{\sqrt{6}}{6} (3 w+1)  \Psi_{(\ell)}\\ &&\qquad+(w-1)\left[ \frac{ \sqrt{6}}{6}\Theta -\frac{ (3 w+1)
   \phi^{\dag}}{6 (w+1)} \right]\Phi_{(\ell)}\;, \label{EqDelEF}\\
\nn&&Z^{\dag}_{(\ell)}=-\frac{2}{3}  \Theta Z_{(\ell)}-2  \phi ^{\dag} \Psi_{(\ell)} +\frac{2 w \left[\Theta ^2+3 (\phi ^{\dag})^2-3 W\right]-3 e^{-\sqrt{\frac{2}{3}} \phi } (3 w+1) \mu}{6 (w+1)}\Delta_{(\ell)}\\ \nn
&& \qquad+ \frac{1}{6\sqrt{6} (w+1)}\left[\sqrt{\frac{8}{3}} (w-1)  \Theta ^2+3 (3 w+1)^2  \mu e^{-\sqrt{\frac{2}{3}}  \phi }+3 (w-1)   (\phi ^{\dag})^2\right.\\
&& \qquad\left.- 6(w-1)  W+3 \sqrt{6} (w+1) W'\right]\Phi_{(\ell)}\;,\\
&&\Phi ^{\dag}_{(\ell)}=\Psi_{(\ell)} -\frac{w   \phi ^{\dag}}{w+1}\Delta_{(\ell)}-\frac{(w-1) \phi ^{\dag}}{\sqrt{6} (w+1)}  \Phi _{(\ell)}\\
\nn&&\Psi ^{\dag}_{(\ell)}=-\Theta  \Psi_{(\ell)} - \phi ^{\dag}Z_{(\ell)}+\frac{1}{ \sqrt{2} (w+1)}\left[  (1-3 w) \mu\,e^{-\sqrt{\frac{2}{3}} \phi }+ \sqrt{6} w\Theta  \phi ^{\dag}\right.\\ &&\qquad\left.+ \sqrt{6} w W'\right]\Delta_{(\ell)}+\frac{1}{12 (w+1)}\left[2 \left(9 w^2-1\right) \mu\,e^{-\sqrt{\frac{2}{3}} \phi }  +6\sqrt{2} (w-1) \Theta  \phi ^{\dag}\right.\\ &&\qquad\left.+6\sqrt{2} (w-1) W' -12\sqrt{3} (w+1) W''\right]\Phi_{(\ell)}\;.
\end{eqnarray}

The most striking difference between the system above and \rf{ScaPertJF} is the structure of the matter fluctuation equation \rf{EqDelEF}. In the Einstein frame the scalar field and its momentum act as a source for the matter fluctuations and influence the dissipation term. Such difference in the behavior of the perturbation in the two frames is due to the change in the structure of the derivative operators.  Also, the structure of the coefficients of the remaining equations is deeply modified, and this will surely induce changes in the behavior of the solution. As we will see in the examples, the difference is particularly evident on large scales, because of the absence of the conserved quantity that characterizes the JF \cite{SantePert}.

%%%%%%%%%%%%%%%%%%%%%%%%%%%%%%%%%%%%%%%%%%%%%%%%%%%%%%%%%
\section{Two examples.}
%%%%%%%%%%%%%%%%%%%%%%%%%%%%%%%%%%%%%%%%%%%%%%%%%%%%%%%%%%
In the following we will explicitly consider two examples, one related to a simple $f(R)$ model in a FLRW background and the other arising from a de Sitter background in an $f(R)$ cosmology.
%%%%%%%%%%%%%%%%%%%%%%%%%%%%%%%%%%%%%%%%%%%%%%%%%%%%%%%%%%
\subsection{The Einstein frame perturbations for $R^{n}$-gravity.}
%%%%%%%%%%%%%%%%%%%%%%%%%%%%%%%%%%%%%%%%%%%%%%%%%%%%%%%%%%
Let us consider the case  $f(R)=\chi R^{n}$, called also sometimes $R^{n}$-gravity, which action reads
\begin{equation}\label{lagrRn}
L=\sqrt{-g}\left[\chi R^{n}+2{\cal L}_{m}(\gimel,{g}^{ab})\right]\;,
\end{equation}
and constitutes the simplest possible example of fourth-order gravity.  We choose this model because its homogeneous and isotropic cosmologies have been studied in detail using the dynamical system approach \cite{ellisbook,Carloni:2004kp}, while the evolution of the large scale cosmological perturbations of a FLRW background has been investigated in  \cite{SantePert,StructForm}  using the covariant gauge invariant approach.  As background solution in the Jordan frame, we will choose the  transient spatially flat solution:
\begin{equation}\label{JFbckgr}
 S=S_0 \left(\frac{t}{t_0}\right)^{\frac{2n}{3(1+w)}}\;.
\end{equation}
Here we will only consider the case $n>\frac{3}{4}(1+\omega)$ for this background, in order to keep the sign of $F=nR^{n-1}$ always positive, consistently with the condition \rf{UpsilonDEF}.
In the Einstein frame \rf{lagrRn} corresponds to the theory
\begin{equation}
\bar{L}=\sqrt{-\bar{g}} \left[ \bar{R} - \bar{g}^{ab} \confnab_{a} \phi \confnab_{b}\phi
- W_0 e^{\sqrt{\frac{2}{3}}\frac{(n-2)}{1-n} \phi} +2e^{-\frac{\phi}{\sqrt{6}} }{\cal L}_{m}(\gimel,\bar{g}^{ab})\right]\;,
\end{equation}
 where $W_0=\frac{1}{2}(\chi)^{\frac{1}{1-n}}n^{\frac{n}{1-n}} (n-1) $.
In turn, \rf{JFbckgr} for $n\neq3/2$ transforms into a solution for the scale factor given by
\begin{equation}
 \bar{S}=\bar{S}_0  \left(\frac{\bar{t}}{\bar{t}_0}\right)^{\frac{n+3 (n-1) w-3}{3 (2 n-3) (w+1)}}\;,
\end{equation}
 where
\begin{equation}
\bar{S}_0 = S_0  \chi^{\frac{1}{6-4 n}} t_0^{\frac{n-1}{2 n-3}} n^{\frac{n}{6-4 n}} (w+1)^{\frac{n-1}{4 n-6}}\left(\frac{3}{2}-n\right)^{\frac{n-1}{2 n-3}}
 \left(\frac{4 n}{3 (w+1)}-1\right)^{\frac{n-1}{6-4
n}} \,,
\end{equation}
 and induces a solution for the scalar field
\begin{equation}
\phi=\phi_0-\frac{1}{(2 n-3)} \sqrt{\frac{3}{2}} \ln\left( \frac{\bar{t}\, {}^{2 (n-1)}}{\chi}\right) \;,
\end{equation}
with
\begin{equation}
\phi_0=\ln \left[3\, n^n  2^{2 n-1} (3-2 n)^{2 (1-n)} (w+1)^{1-n} \left(\frac{8 n}{w+1}-6\right) \left(\frac{4 n}{3 w+3}-1\right)^n\right]\,.
   \end{equation}
If $n=3/2$, instead, one obtains
\begin{equation}
 \bar{S}=\bar{S}_0 e^{-\frac{(n+3 (n-1) w-3) \bar{t}+2 n \bar{t}_0}{3 \chi \sqrt{3n[4 n-3 (w+1)]} }}\;,
\end{equation}
where now
\begin{equation}
\bar{S}_0= 2^{n-1} n^{n/2} S_0 \left(\frac{4 n}{3}-w-1\right)^{\frac{n-1}{2}} (w+1)^{1-n} \sqrt{\chi } \;,
\end{equation}
and induces a solution for the scalar field
\begin{equation}
\phi=\phi_0-\frac{2 (n-1) (w+1) \bar{t}}{\chi \sqrt{3n[4 n-3 (w+1)]} } \;,
\end{equation}
with
\begin{equation}
\phi_0=\sqrt{\frac{3}{2}} \ln \left[4^{n-1} n  \chi  \left(\frac{n}{w+1}\right)^{n-1} \left(\frac{4 n}{3 (w+1)}-1\right)^{n-1}\right]\,.
\end{equation}
This last case is particularly interesting because it explicitly shows how a non accelerating background is in fact transformed into a de Sitter solution via a conformal transformation.

\begin{figure}[htbp]
\subfigure[The time evolution of the long wavelength density fluctuations in the Jordan frame for $n=1.1$.]{\includegraphics[scale=0.5]{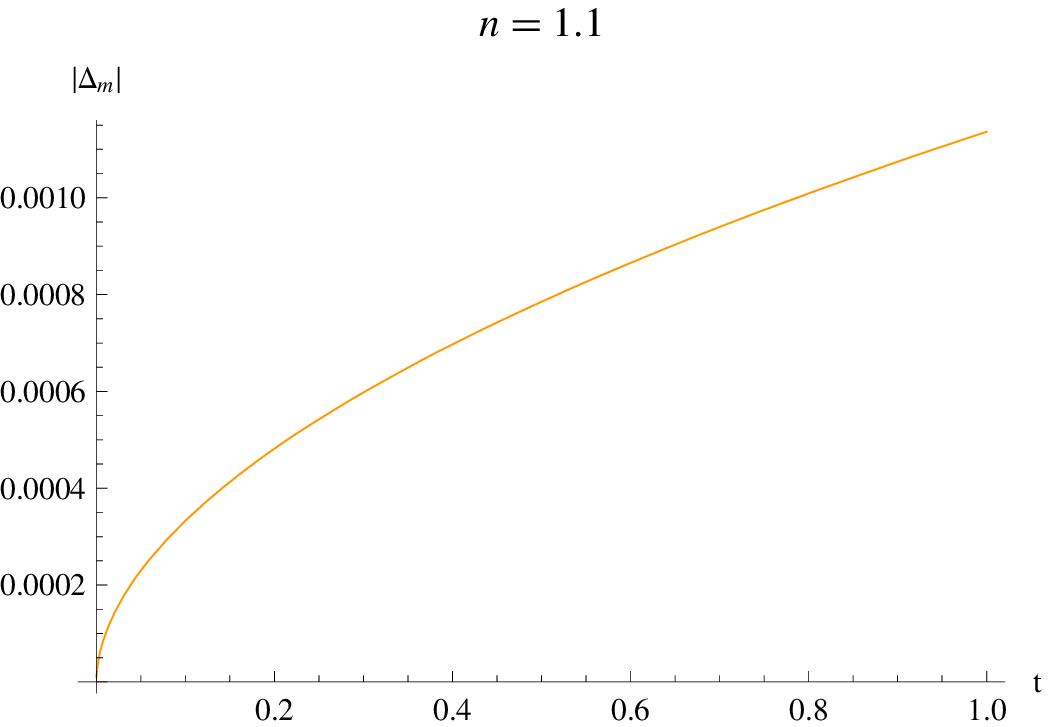}}
\subfigure[The time evolution of the long wavelength density fluctuations in the Einstein frame for $n=1.1$.]{\includegraphics[scale=0.5]{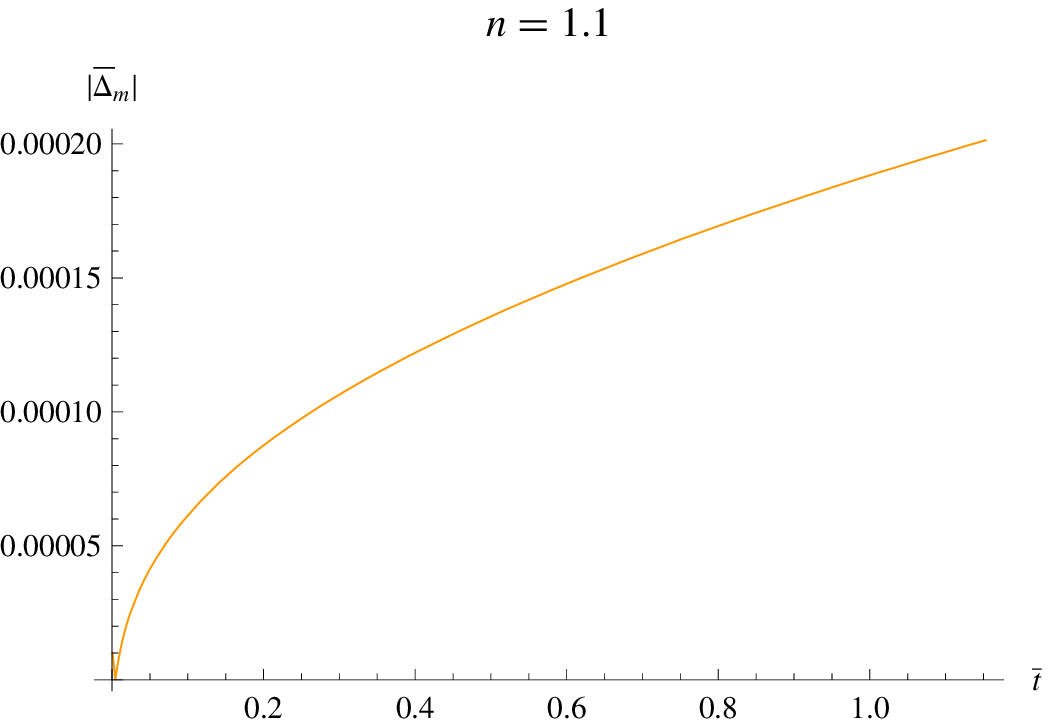}}
\subfigure[The time evolution of the long wavelength density fluctuations in the Einstein frame for $n=1.2$.]{\includegraphics[scale=0.5]{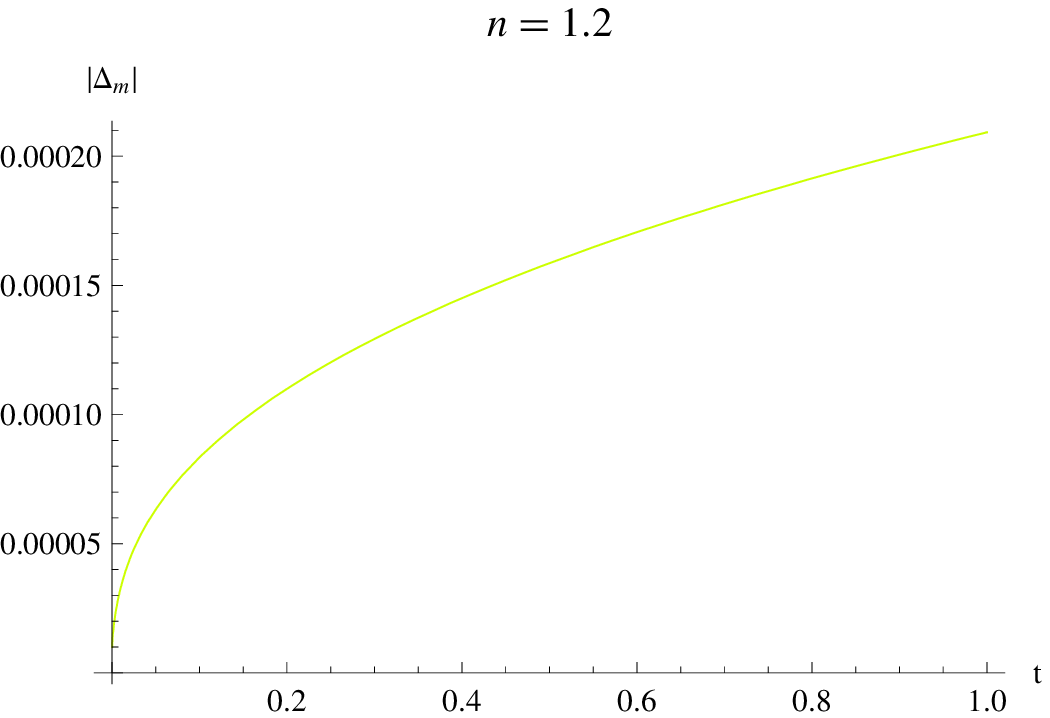}}
\subfigure[The time evolution of the long wavelength density fluctuations in the Einstein frame for $n=1.2$.]{\includegraphics[scale=0.5]{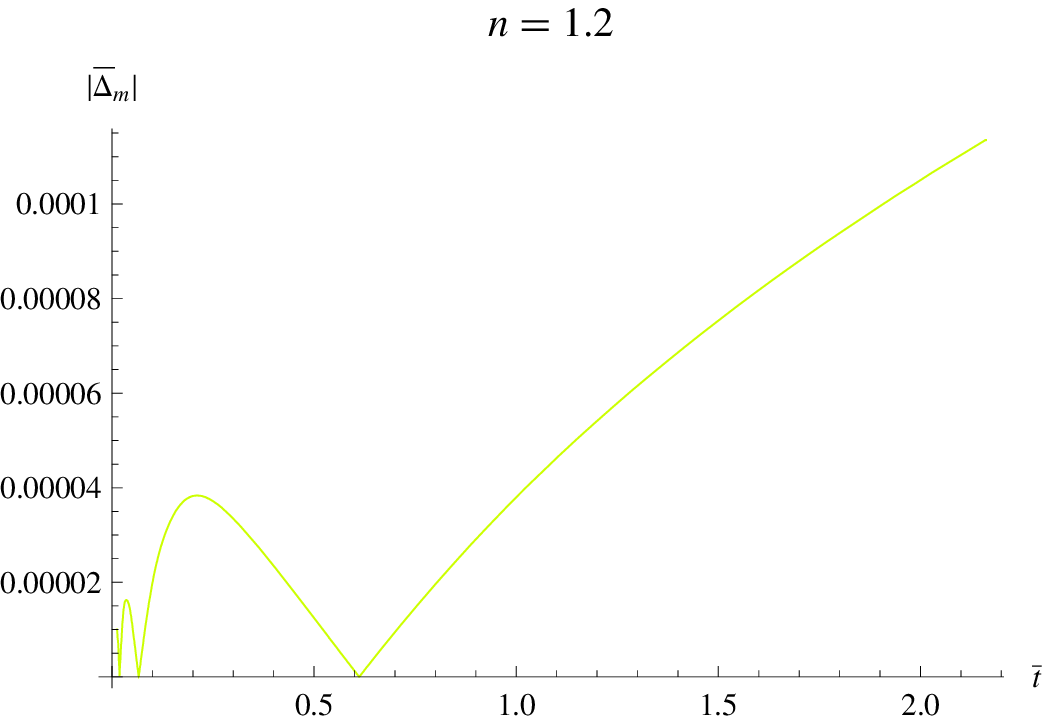}}
\caption{Comparison of the time evolution of the long wavelength density fluctuations in the Jordan frame  and the Einstein frame  in the case of $R^n$-gravity and dust. Note that for $n\rightarrow1$ the two solutions tend to converge to  the GR solution. This is due to the fact that in this limit in both frames the equations tend to the GR ones.}\label{FigPert1}
\end{figure}

\begin{figure}[htbp]
\subfigure[The time evolution of the long wavelength density fluctuations in the Einstein frame for $n=1.3$.]{\includegraphics[scale=0.5]{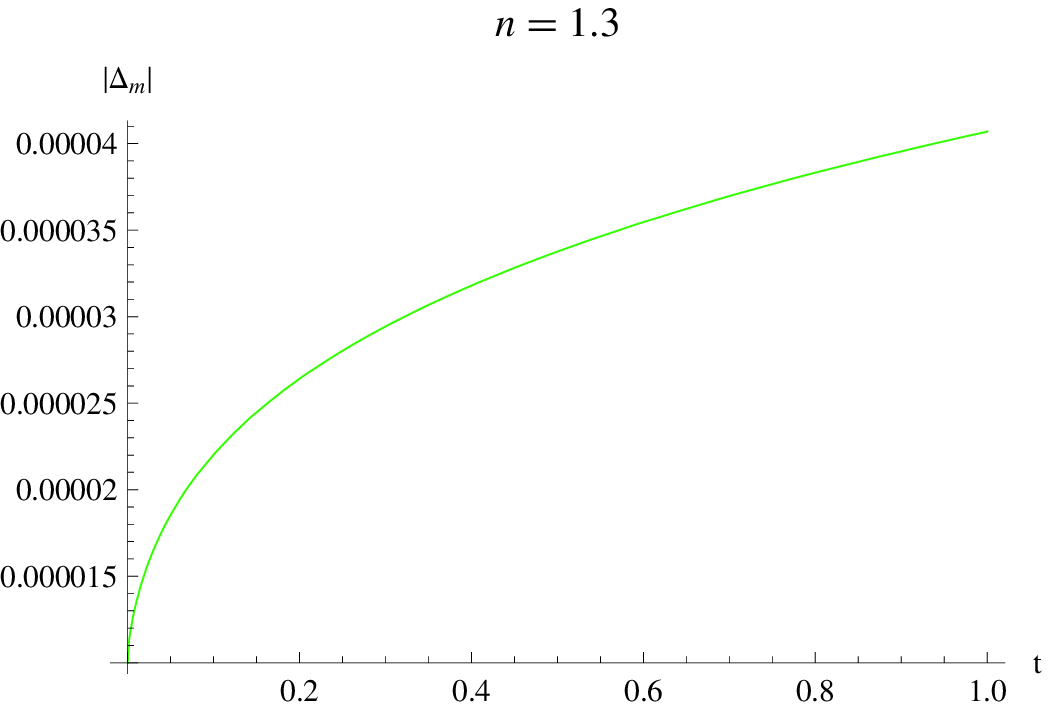}}
\subfigure[The time evolution of the long wavelength density fluctuations in the Einstein frame for $n=1.3$.]{\includegraphics[scale=0.5]{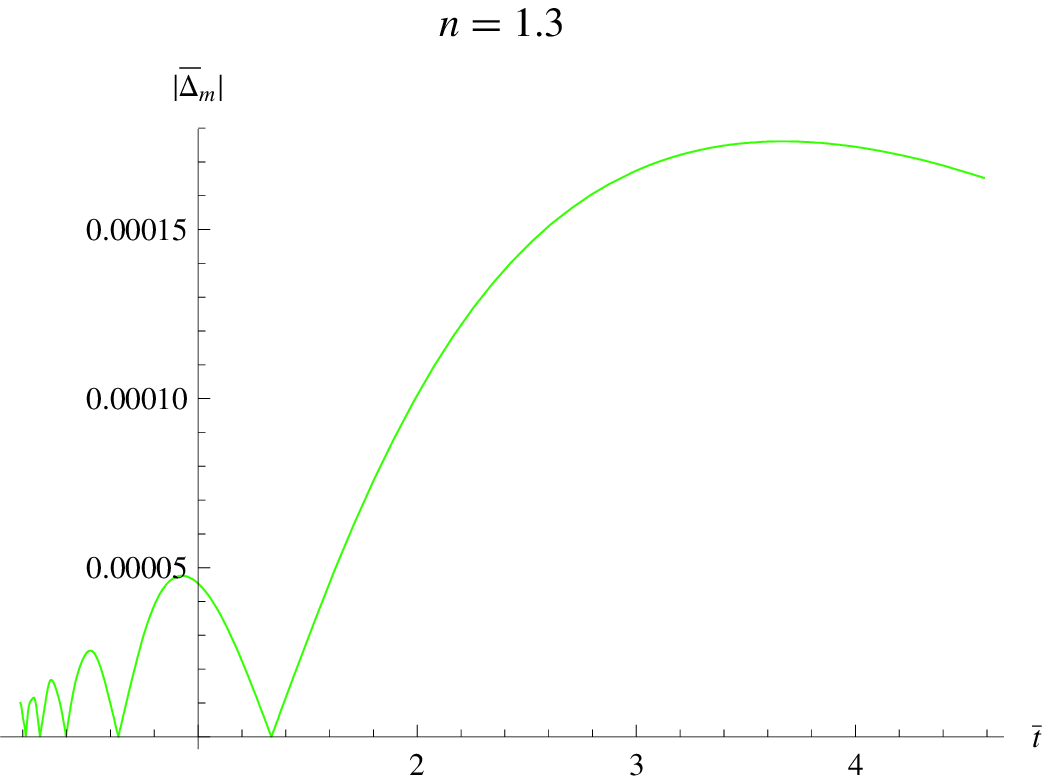}}
\subfigure[The time evolution of the long wavelength density fluctuations in the Einstein frame for $n=1.4$.]{\includegraphics[scale=0.56]{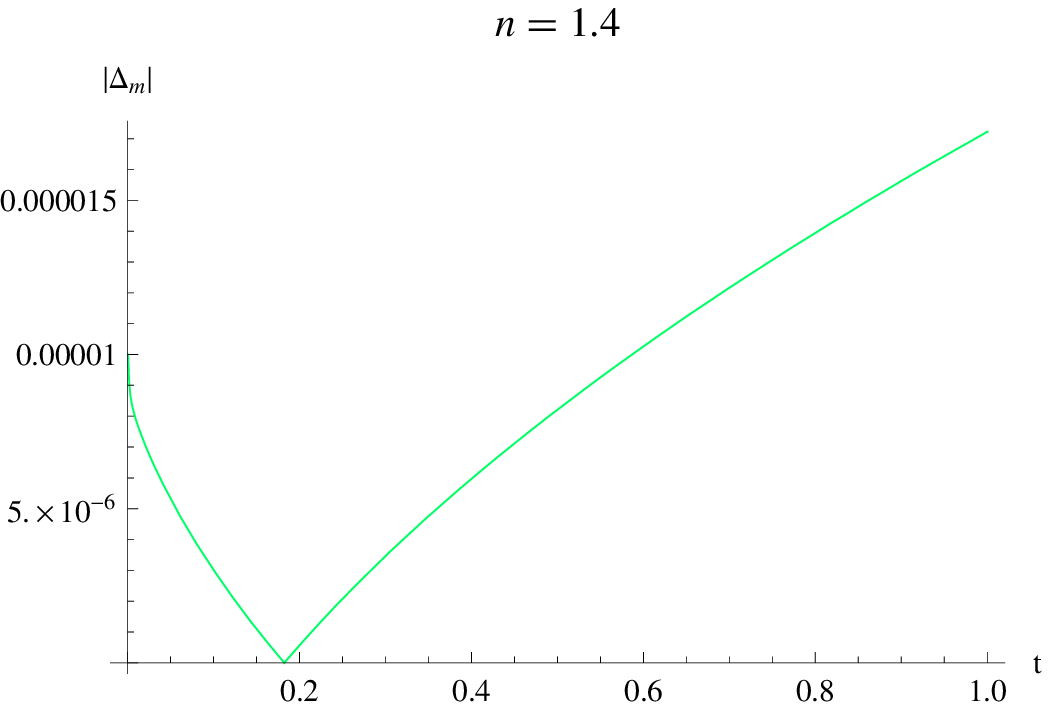}}
\subfigure[The time evolution of the long wavelength density fluctuations in the Einstein frame for $n=1.4$.]{\includegraphics[scale=0.56]{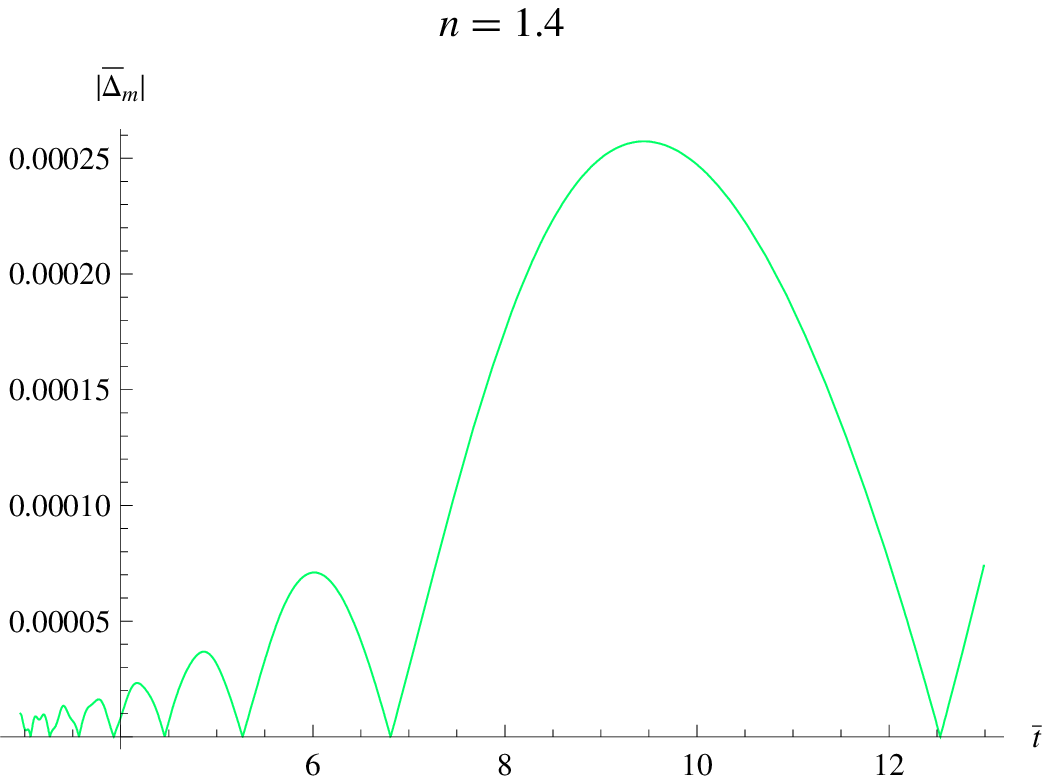}}
\caption{Comparison of the time evolution of the long wavelength density fluctuations in the Jordan frame  and the Einstein frame  in the case of $R^n$-gravity and dust. }\label{FigPert2}
\end{figure}

Introducing these solutions in the two scalar perturbation systems \rf{ScaPertJF} and \rf{ScaPertEF}, we are able to calculate numerically the evolution of the scalar fluctuations in the two frames.  The results we obtained for the long wavelength dust ($\omega=0$) fluctuations, with different values of the parameter $n$, are shown in Figs.~\ref{FigPert1} and  \ref{FigPert2}. It is clear from these that, as expected from the general equations, the behavior of the scalar perturbations differs in the two conformal frames. In particular one can notice that the growth rate of the fluctuation becomes more and more different when $n$ increases. For example, for $n=1.4$ the perturbations in the Jordan frame decay, while they still grow in the Einstein frame. Moreover, the JF  perturbations on the several scales evolve clearly with a power law behavior, while the EF ones  oscillate visibly, as expected form the general considerations in the previous section. Finally one can see that for $n\rightarrow1$ the differences in the matter fluctuations in the two frames tend to disappear. This happens because in the Jordan frame the  fourth-order terms, being multiplied by the $n-1$ factor, become more and more suppressed  and the corresponding theory tends to ordinary General Relativity. A similar phenomenon happens in the Einstein frame:  for $n\rightarrow 1$ the scalar field is related to $R^{n-1}$  and tends to a constant while its potential tends to zero  so that the theory corresponds once more to pure Einsteinian gravity.
%%%%%%%%%%%%%%%%%%%%%%%%%%%%%%%%%%%%%%%%%%%%%
%%%%%%%%%%%%%%%%%%%%%%%%%%%%%%%%%%%%%%%%%%%%%%%%%%%%
%%%%%%%%%%%%%%%%%%%%%%%%%%%%%%%%%%%%%%%%%%%%%
%%%%%%%%%%%%%%%%%%%%%%%%%%%%%%%%%%%%%%%%%%%%%%%%%%%%

%%%%%%%%%%%%%%%%%%%%%%%%%%%%%%%%%%%%%%%%%%%%%%%%%%%%%%%%%%
\subsection{Perturbations of the de Sitter spacetime in $f(R)$-gravity.}
%%%%%%%%%%%%%%%%%%%%%%%%%%%%%%%%%%%%%%%%%%%%%%%%%%%%%%%%%%
In this section we will compare the properties of the Jordan and of the Einstein frames of $f(R)$ cosmological models characterized by a de Sitter background. The presence of such background(s) in $f(R)$-gravity is one of the most important features of these theories because it has the potential to model both inflation and dark energy \cite{Nojiri:2007as,Cognola:2007zu}. In fact,  it has been proven that a viable $f(R)$-gravity model unifying inflation  and late time acceleration in the form of double de Sitter solution can be always constructed numerically \cite{Cognola:2008zp}.  As we will see, however, such backgrounds are not suitable for structure formation, because matter, even if present in non negligible quantities, is dissipated very quickly. Notwithstanding this physical issue, the peculiar properties of this metric allow us to  go deeper in understanding the difference between the two frames.
%%%%%%%%%%%%%%%%%%%%%%%%%%%%%%%%%%%%%%%%%%%%%%%%%%%%%%%%%%
\subsubsection{Perturbations of the vacuum de Sitter spacetime in $f(R)$-gravity.}
%%%%%%%%%%%%%%%%%%%%%%%%%%%%%%%%%%%%%%%%%%%%%%%%%%%%%%%%%%
Let us consider a Universe in which the background is given by a de Sitter spacetime characterized by a scale factor $S=S_0 e^{\beta t}$ and vacuum ($\mu=0$). Substituting in the cosmological equations it is easy to show that $\beta$ has to satisfy the equation
\begin{equation}
18 \beta ^2 F_0-f_0=0\,,
\end{equation}
where $F_0=F(R_0)$, $f_0=f(R_0)$ and $R_0=12\beta^{2}$. Let us now consider a perturbation of this spacetime in which a fluid, constituted for instance by standard matter, is present. According to what has been said in the previous section on this model, this fluid will be described by first-order quantities. We will also assume that the fluid is actually barotropic in its rest frame i.e. its equation of state is $p=\omega \mu$. Choosing a set of observers comoving with it\footnote{Since the definition of the fluid flow $u_a$ is made at the level of the perturbed Universe this choice is legitimate. In addition to that, the de Sitter solution is frame invariant so any choice of frame in the background is equivalent.}, the harmonically decomposed perturbation equations reduce to
\begin{eqnarray}
&&\dot{\Delta}^{(\ell)}=-3 \beta  (\omega +1) \Delta^{(\ell)}\,, \\
&&\ddot{\mathcal{R}}^{(\ell)}+3 \beta
   \dot{\mathcal{R}}^{(\ell)}+\left[\frac{ e^{-2 t \beta } \ell
   ^2}{S_0^2}-12 \beta^2+\frac{4 f_0}{F_0}+\frac{2
   F_0}{F_0'}\right]\mathcal{R}^{(\ell)} -\frac{(1-3 \omega ) \Delta^{(\ell)} }{3 F_0'}\,.
\end{eqnarray}
In this system the equation for $\Delta$ is scale invariant and, as expected, matter perturbations are exponentially suppressed with a time constant which depends on $\omega$ and the time constant of the de Sitter solution. The curvature perturbations, instead, are governed by a second-order equation which is forced by the matter term.

In the long wavelength limit $\ell=0$ the above equations yield the general solutions
\begin{eqnarray}
&&\Delta=\Delta _0 \,e^{-3 t  \beta  (1+\omega )} =\Delta _0 \left(\frac{S}{S_0}\right)^{-3 (\omega  +1 )}, \\
&& \mathcal{R}=\mathcal{R}_{0,1}\,e^{-3 t \beta  (1+\omega )} + \mathcal{R}_{0,2}\, e^{t \alpha _+}+\mathcal{R}_{0,3}\,e^{t \alpha _-},
\end{eqnarray}
where
\begin{equation}
\alpha_\pm=-3 \beta\pm\sqrt{25 \beta ^2-\frac{4 F_0}{3 F_0'}}\,.
\end{equation}
Here $\mathcal{R}_{0,i}$  and $\Delta _0$ are constants of integration and we have dropped the apex ``${(0)}$" to make the notation lighter. It is plain from this solution that, in a de Sitter background, standard matter is clearly made homogeneous, but this is not the case for the perturbation of the Ricci curvature. If one considers $\mathcal{R}$ as representing the scalar gravitational waves normally associated to this type of theories, one can see that, depending on the form of the function $f$, this kind of perturbation is able to grow. In addition, if we imagine our $f(R)$-model to be a (classic) inflationary one, we can see that the analysis of these scalar waves would constitute a direct and purely classical test of the nature of the gravitational interaction, based on the gravitational wave relic of the inflationary era. This is a result worth exploring.

In the Einstein frame, the background $S$ corresponds to another de Sitter background, given by
\begin{equation}
\bar{S}=\bar{S}_0 e^{\bar{t}\zeta}\,,\qquad \zeta=\frac{ \beta }{ F_0}=\beta\,e^{-\sqrt{2/3} \phi_0}\,,\qquad \phi=\phi_0=\sqrt{\frac{3}{2}} \log \left[ F_0\right]\,,
\end{equation}
where $\bar{S}_0=S_0\sqrt{F_0}$. The perturbation equations become
\begin{eqnarray}
&&\bar{\Delta}^{\dag}_{(\ell)} =\frac{\beta}{3}e^{-\sqrt{2/3} \phi_0} \bar{ \Delta}_{(\ell)} \,,\\
&&\bar{\Phi}^{\dag\dag}_{(\ell)} = -\beta \,e^{-\sqrt{2/3} \phi_0} \bar{\Phi}^{\dag}_{(\ell)} -\frac{\ell^2}{ \bar{S}_0^2 } \bar{\Phi}_{(\ell)} \,e^{-2\bar{t}\zeta-\sqrt{2/3} \phi_0}\,.
\end{eqnarray}
It is clear that, also in the Einstein frame, standard matter perturbations are scale invariant, and they are dissipated by the expansion, but in this frame they do not act as a source of the perturbation of the scalar degree of freedom. In fact the two equations above are decoupled and can be solved exactly. In the long wavelength limit $\ell=0$, one has
\begin{eqnarray}
&&\bar{\Delta}=\bar{\Delta}_0\, e^{\bar{t} \zeta}=\bar{\Delta}_0 \left(\frac{\bar{S}}{\bar{S}_0}\right)^{1/3}\,,\\
&&\bar{\Phi} =\bar{\Phi}_{0,1}-\,\frac{\bar{\Phi}_{0,2}}{\beta }\,e^{-\bar{t}\zeta+\sqrt{2/3} \phi_0}\,.
\end{eqnarray}
where we have dropped again the apex ``${(0)}$". Note how, because of the modification in the conservation equations, the evolution of the density perturbation in the Einstein frame does not depend on the barotropic factor of the matter fluid. Also the behavior of $\Phi$ can be very different from the one of $ \mathcal{R}$.  For $F>0$, for example, the first quantity is damped and converges exponentially to a constant value, while the latter one can exhibit a very different behavior.

The form of the exponents of the modes the solutions above for some popular models of $f(R)$-gravity \cite{shosho,Nojiri:2007as,Cognola:2007zu} are given in Table \ref{dSTab}.

\begin{table}[htdp]
\caption{Some of the values of $\beta$  and the exponents of the modes of the scalar fluctuation solutions for various popular $f(R)$-gravity models in pure de Sitter backgrounds in the Jordan and in the Einstein frames. For the more complex forms of $f(R)$ the implicit equation to be solved in order to find the parameters have been given. Of special interest are the models $f(R)=\frac{R^m+\chi}{1+\xi R^n}$ and their generalizations, which can provide a unique theoretical framework for early time inflation and late time acceleration \cite{Nojiri:2007as,Cognola:2007zu} (the first unified models of this type were proposed in \cite{Nojiri:2003ft,Capozziello:2006dj}).}\label{dSTab}
\begin{center}
\scriptsize{\begin{tabular}{cccc}
\hline\hline
 $f(R)$ & $\beta$ & $\alpha _\pm$
 \\\hline
$R+\chi R^n$ &  $\left(2^{2 n-1} 3^{n-1} \alpha -3^n 4^{n-1} n \chi \right)^{\frac{1}{2-2 n}}$ & $-3 \beta \pm A$
\\&&&\\
$\exp(q R)$ &  $\pm\frac{1}{3\sqrt{2q}}$ & $\pm\frac{2}{3}\sqrt{\frac{2}{q}}$

\\&&&\\
$\frac{R^m+\chi}{1+\xi R^n}$ &
$\frac{3^{n+1} 4^n n \beta ^{2 n} \left(12^m \xi  \beta ^{2 m}+1\right)-\left(3^m \left(3 4^m m+2^{2 m+1}\right) \xi
   \beta ^{2 m}+2\right) \left(12^n \beta ^{2 n}+\chi \right)}{12 \left(12^m \xi  \beta ^{2 m}+1\right)^2}=0$ &$-3 \beta \pm \sqrt{25\beta ^2+B }$
 \\&&&\\
$R+\chi+\frac{\chi }{\alpha  \left[(R \beta -1)^{2
   n+1}+1\right]+1}$& $\beta ^2-\frac{6 \beta ^2 \xi ^2 \chi ^2 (2 n+1) \left(12 \beta ^2 \xi -1\right)^{2 n}}{\left(2 \left(12
   \xi -1\right)^{2 n+1}+\xi  \chi +2\right)^2}+\frac{\chi}{6} -\frac{\xi  \chi ^2}{12 \left(12 \beta ^2
   \xi -1\right)^{2 n+1}+\xi  \chi +2}=0$& $-3 \beta \pm \sqrt{25 \beta ^2-D}$
 \\&&&\\\hline
\multicolumn4c{$A=\sqrt{\frac{25 n^2+7n-32}{3n (n-1)}}\left(2^{2 n+3}-3 n 4^n\right)^{\frac{1}{2-2 n}} \chi ^{\frac{1}{2-2 n}}$}\\\multicolumn4c{$B=\frac{16 \left(12^m \xi  \beta ^{2 m}+1\right) \left(12^n n \beta ^{2 n} \left(12^m \xi  \beta ^{2 m}+1\right)-12^m
   m \beta ^{2 m} \xi  \left(12^n \beta ^{2 n}+\chi \right)\right)}{12^m m \xi  \left(12^m (m+1) \xi  \beta ^{2
   m}-m+1\right) \left(12^n \beta ^{2 n}+\chi \right) \beta ^{2 m}+12^n (n-1) n \left(12^m \xi  \beta ^{2
   m}+1\right)^2 \beta ^{2 n}-2^{2 m+2 n+1} 3^{m+n} m n \xi  \left(12^m \xi  \beta ^{2 m}+1\right) \beta ^{2 (m+n)}}$}\\\multicolumn4c{$C=\frac{2 \left(24 n \beta ^2 \xi  \chi  \left(144 \xi  \beta ^4+1\right)^{-n-1}+1\right)
   \left(144 \xi  \beta ^4+1\right)^{n+2}}{3 n \xi  \left(144 (2 n+1) \beta ^4 \xi -1\right) \chi }$}\\\multicolumn4c{$D=\frac{\left(12 \beta ^2 \xi -1\right)^{1-2 n} \left(2 \left(12 \beta ^2 \xi -1\right)^{2 n+1}+\xi  \chi +2\right)^3}{3 (2 n+1) \xi ^3 \chi ^2 \left(2 (n+1) \left(12 \beta ^2 \xi -1\right)^{2
   n+1}-n (\xi  \chi +2)\right)}
   \left(1-\frac{2 (2 n+1) \xi ^2 \left(12 \beta ^2 \xi -1\right)^{2 n} \chi ^2}{\left(2 \left(12 \beta ^2 \xi
   -1\right)^{2 n+1}+\xi  \chi +2\right)^2}\right)$}\\ \hline\hline
\end{tabular}}\\
\vspace{0.5cm}
\scriptsize{\begin{tabular}{cccc}
 \hline\hline $f(R)$ &  $\zeta$\\\hline
$R+\chi R^n$ &$\frac{2^{\frac{1}{n-1}} (2-3 n) \left(3^{n-1} \left(2^{2 n+1}-3 4^n n\right) \chi \right)^{\frac{1}{2-2 n}}}{2-2 n}$\\&&&\\
$\exp(q R)$&$\pm\frac{1}{3 \sqrt{2} e^{2/3} q^{3/2}}$\\&&&\\
$\frac{R^m+\chi}{1+\xi R^n}$ &
 $\frac{12 \beta ^3 \left(12^m \xi  \beta ^{2 m}+1\right)^2}{12^n n \beta ^{2 n} \left(12^m \xi  \beta ^{2
   m}+1\right)-12^m m \beta ^{2 m} \xi  \left(12^n \beta ^{2 n}+\chi \right)}$\\&&&\\				 
$R-\chi\left[\left(\frac{1}{\xi
   R^2+1}\right)^n-1\right] $&$\frac{\beta }{24 n \beta ^2 \xi  \chi  \left(144 \xi  \beta ^4+1\right)^{-n-1}+1}$\\&&&\\
$R+\chi+\frac{\chi }{\alpha  \left[(R \beta -1)^{2
   n+1}+1\right]+1}$& $\frac{\beta }{1-\frac{2 (2 n+1) \xi ^2 \left(12 \beta ^2 \xi -1\right)^{2 n} \chi ^2}{\left(2 \left(12 \beta ^2 \xi-1\right)^{2 n+1}+\xi  \chi +2\right)^2}}$\\ \hline\hline
\end{tabular}}
\end{center}
\label{default}
\end{table}

%%%%%%%%%%%%%%%%%%%%%%%%%%%%%%%%%%%%%%%%%%%%%%%%%%%%%%%%%%
\subsubsection{Perturbations of the non-vacuum de Sitter spacetime in $f(R)$-gravity.}
%%%%%%%%%%%%%%%%%%%%%%%%%%%%%%%%%%%%%%%%%%%%%%%%%%%%%%%%%%
The background solutions we have considered so far are purely-vacuum solutions. This is due to the fact that the cosmological equations do not seem to be compatible with the de Sitter evolution in presence of matter. However, one can find special equations of state for which a de Sitter solution can exist within $f(R)$-gravity which is actually compatible with a non zero (although constant) energy density.

Let us consider an homogeneous and isotropic cosmology with a fluid with equation of state $p=\gamma \mu +\xi$ where $\gamma$ is a barotropic factor, $\mu$ is (as before) the matter energy density and $\xi=\frac{1}{2} (\gamma +1) \left[f_0-18 \beta ^2 F_0\right]$\footnote{Note that this fluid is physical only as far as $\mu$ is different from zero, because in the vacuum case one would have a pressure which is not associated to any energy density. In the following we will always assume $\mu\neq0$.}.  Then a generic de Sitter spacetime $S=S_0 e^{\beta t}$ is a solution for this cosmology provided that $\mu=\mu_*$ is constant and $\mu_*=- \frac {\xi}{1+\gamma}$.
If one derives the scalar perturbation equations in the Jordan Frame, one obtains, after harmonic decomposition,
\begin{eqnarray}
&&\dot{\Delta}^{\ell}=-3 \beta  (\gamma +1) \Delta^{\ell}\,, \\
&&\ddot{\mathcal{R}}^{\ell}+3 \beta
   \dot{\mathcal{R}}^{\ell}+\left[\frac{ e^{-2 t \beta } \ell
   ^2}{S_0^2}-12 \beta^2+\frac{4 f_0}{F_0}+\frac{2
   F_0}{F'_0}\right]\mathcal{R}^{\ell} -\frac{(1-3 \gamma ) \Delta^{\ell} }{3 F'_0}\,.
\end{eqnarray}
In this system, as expected, the matter perturbations are exponentially suppressed with a time constant which depends on $\gamma$ and on the time constant of the de Sitter solution, while the curvature perturbations are forced by the matter term.
In the long wavelength limit $\ell=0$ the above equations yield the general solutions
\begin{eqnarray}\label{SolJFdS+m}
&&\Delta=\Delta _0 e^{-3 t (\omega  \beta +\beta )}\;, \\
&& \mathcal{R}=e^{-3 t (\omega  \beta +\beta )} c_1+e^{t \alpha _+} c_2+e^{t \alpha _-} c_3\;,
\end{eqnarray}
where
\begin{equation}
\alpha_\pm=-3 \beta\pm\sqrt{57 \beta ^2-\frac{8 f_0}{3 F_0}-\frac{4 F_0}{3 F'_0}}\;,
\end{equation}
and $c_i$ are constants of integration.

In the Einstein frame the background $S$ transforms again into the de Sitter background seen in the previous case plus a constant scalar field, however the presence of matter in both the background and the perturbed Universe makes the differences between the two frames even more evident.
The perturbation equations in the Einstein frame read
\begin{eqnarray}
&&\bar{\Delta}^{\dag}_{\ell} =-(\gamma+1) \beta \bar{\Delta}_{\ell}\,  e^{-\sqrt{2/3} \phi_0}-\sqrt{\frac{2}{3}}  \Phi ^{\dag}_{\ell}\,,\\
&& \Phi^{\dag\dag}_{\ell}= -\beta\Phi^{\dag}_{\ell}e^{-\sqrt{2/3} \phi_0}+
\left(-\frac{ \ell^2}{\bar{S}_0^2 }e^{-4\bar{t}\zeta-\sqrt{2/3} \phi_0}+\frac{\xi (3 \gamma-1)}{3 (1+\gamma)} e^{-\sqrt{2/3} \phi_0}\right)\Phi_{\ell}\nn\\ && \qquad +\frac{\xi}{\sqrt{6}}  \frac{(1-3 \gamma)}{1+\gamma}  \bar{\Delta}_{\ell}\, e^{-\sqrt{2/3} \phi_0}\,.
\end{eqnarray}
Comparing with \rf{SolJFdS+m}, one finds that the equation describing the matter perturbations does not decouple, so that the interaction between the scalar degrees of freedom and matter is more pronounced.  The solution of this system reads
\begin{eqnarray}
&& \bar{\Delta}=\bar{\Delta}_{0,1}\,e^{\delta_1\bar{t}}+\bar{\Delta}_{0,2}\,e^{\delta_2\bar{t}}+\bar{\Delta}_{0,3}\,e^{\delta_3\bar{t}}\;,\\
&& \Phi=\Phi_{0,1}\,e^{\alpha_1\bar{t}}+\Phi_{0,2}\,e^{\alpha_2\bar{t}}+\Phi_{0,3}\,
e^{\alpha_3\bar{t}},
\end{eqnarray}
where $\Phi_{0,i}$ and $\bar{\Delta}_{0,i}$ are integration constants, while $\alpha_i$ and $\delta_i$ are the solutions of the equations
\begin{eqnarray}
\nn&& \left[e^{\bar{t} \delta -\lambda  \phi _0} \beta  (\gamma +1) +e^{\bar{t} \delta } \delta\right]  \Delta _0+e^{\bar{t} \alpha} \alpha  \lambda  \Phi _0=0\,,\\
\nn&&\left\{e^{\bar{t} \alpha } \alpha ^2 -\left[-e^{\bar{t} \alpha -\lambda  \phi _0} \alpha
   \beta -\frac{1}{6} e^{\bar{t} \alpha -\lambda  \phi _0} (\gamma +1) (3 \gamma -1) \left(18 e^{\lambda  \phi _0} \beta
   ^2-f_0\right)\right]\right\} \Phi _0\\&&-\frac{1}{4} e^{\bar{t} \delta-\lambda  \phi _0} (\gamma +1) (3
   \gamma -1) \lambda  \left(18 e^{\lambda  \phi _0} \beta ^2-f_0\right) \Delta _0=0\,.
\end{eqnarray}
It seems clear that  the presence of matter at the background level enhances the  differences between the equations in the two frames, JF and EF. This obviously propagates to their solutions too. As an example, the matter fluctuations in the Einstein frame contain now three modes while the ones in the Jordan frame contain only one.

%%%%%%%%%%%%%%%%%%%%%%%%%%%%%%%%%%%%%%%%%%%%%%%%%%%%%%%%%
\section{Discussion and Conclusions}
%%%%%%%%%%%%%%%%%%%%%%%%%%%%%%%%%%%%%%%%%%%%%%%%%%%%%%%%%%
In this paper we have used the 1+3 covariant approach and the CoGI approach to investigate the physics of conformal transformations. We have shown that, what is ordinarily called a conformal transformation is in fact the combination of two different transformations: the usual geometrical transformation of the metric, first, and a subsequent redefinition of the fields in the theory.
The two transformations are independent from each other and each one has its own specific meaning. In particular,  the geometrical conformal transformation can be seen as the passage to an observer, called conformal, which is non-inertial, e.g., the conformal observer possesses an acceleration with respect to the reference observer in JF and perceives a warped spatial metric. In operational terms, this implies that the rods and clocks of this observer have a rate and length which depend on the spacetime coordinates. In performing this transformation, that is a (relatively) simple change of observers, no change in the model occurs.

A real change can appear only in the second transformation, when the fields are redefined. This redefinition corresponds, operatively, to impose  to the conformal observer to be inertial and to assuming that all the non-inertial effects the conformal observer perceives are, in fact, due to a new interaction. This realization physically clarifies both the origin and nature of the scalar field in the Einstein frame: this field is not a new form of matter energy, but just a kinematic effect, conceptually not dissimilar from a non-inertial force in classical mechanics. In some cases, like the scalar tensor  gravity one, such change is masked by the fact that there are ways to redefine a scalar field in the action which do not change any other aspects of the theory. In others, like $f(R)$-gravity, the situation is more delicate and, as we have seen, the effect of this transformation becomes more evident.

The new scalar field is precisely the key to the important simplification that  non-standard gravitational models undergo upon being conformally transformed. In particular, when one transforms a specific theory of gravity, the conformal transformation is chosen in such a way that the kinematic effects compensate the non-Einsteinian contribution to the theory. As a consequence, the transformation reduces non-standard gravity to standard General Relativity plus a scalar field. It is interesting to note that this new field turns out to be non-minimally coupled to matter only if a non-minimal coupling is already present in the theory. In other words, the conformal transformation does not generate non-minimal coupling between  the scalar field and  standard matter.
 
In this paper we have explicitly derived the 1+3 kinematical and thermodynamical variables and this has given us an idea of the general action of conformal transformations on a cosmological model. In particular we have seen that, as expected, the physics in the Einstein frame can have characteristics which are completely different from the ones arising in the Jordan frame. Even if some of the geometrical properties of the cosmology are preserved (homogeneous and isotropic Universes are mapped into homogeneous and isotropic universes), its behavior can be very different. As we have seen explicitly, it can even happen that decelerating cosmologies are mapped in accelerated ones.

These differences become even more pronounced when we consider first-order perturbations. In particular, from the 1+3 equations it is quite clear that the structure of first-order vector and tensor perturbations are not affected by the conformal transformation, but the same cannot be said of the scalar perturbations, which include the matter density fluctuations. The behavior of these quantities appears to be very different in the two frames, not only in terms of the growth rate, but also concerning general evolutionary features, as the presence or absence of oscillations, and so on. 

In recent years, the issue has been raised that the difference we have encountered between the Jordan and the Einstein frame are only apparent, because they do not take in account that we usually perform measurement comparing homogeneous physical quantities and in these comparisons the conformal factor can be cancelled out \cite{Flanagan:2004bz,Sotiriou:2007zu}. Although this is certainly true for (some specific!) local measurements, it is also true that a  theory of measurement in GR has not yet been formulated \cite{Sachs:1967,Boniolo:2000wq}, and one should refer to this theory to determine if the differences in the cosmologies of the conformal frames are apparent or not. The development of such theory and the analysis of its consequences is well beyond the scope of this paper and it will be pursued elsewhere.

We would like to conclude saying that the results above show clearly that our analysis  provides a set of very efficient tools to perform a thorough analysis of conformally related cosmological models. Even if we have used these tools to probe the differences between the two conformal frames, the transformations equations can be also used  to translate results obtained in one of the frames to another, or even to define new forms of the conformal factor specifically tailored to analyze different aspects of the theories considered. The mathematical structure of the 1+3 formalism guarantees that this is possible also when one introduces approximations. Hence, seen through the 1+3 approach, the conformal transformation becomes a powerful tool, able to help in the analysis of complicated alternative gravity models in ways so far unexpected, which deserves further investigation.

\appendix
%%%%%%%%%%%%%%%%%%%%%%%%%%%%%%%%%%%%%%%%%%%%%%%%%%%%%%%%%
\section{The 1+3 Equations and their form in the Conformal frame.}\label{AppA}
%%%%%%%%%%%%%%%%%%%%%%%%%%%%%%%%%%%%%%%%%%%%%%%%%%%%%%%%%%
In this appendix we will list explicitly all the 1+3 covariant equations in the different frames.  Here $\mu$ and $p$ represent the total energy density and pressure, respectively, that one would define when the field equations are in the form $G_{ab}=T^{tot}_{ab}$.
%%%%%%%%%%%%%%%%%%%%%%%%%%%%%%%%%%%%%%%%%%%%%%%%%%
\subsubsection*{The general 1+3 equations.}
%%%%%%%%%%%%%%%%%%%%%%%%%%%%%%%%%%%%%%%%%%%%%%%%%%%
\noindent Expansion propagation (generalized Raychaudhuri equation):
\begin{eqnarray}\label{1+3eqRay}
&&{\dot{\Theta}+{\textstyle\frac{1}{3}}\Theta^2 -\tilde{\nabla}^a
a_{{a}}- a_{{a}}
a^{{a}}+2\sigma_{{{a}}{{b}}}
\sigma^{{{a}}{{b}}} -2\omega_{{a}}\omega^{{a}}+\frac{1}{2}}(\mu +
3p) =0\;.\end{eqnarray}
Vorticity propagation:
\begin{equation}\label{1+3Vor}
\dot{\omega}_{\langle {{a}}\rangle }
+{\textstyle\frac{2}{3}}\Theta\omega_{{a}} -{\textstyle\frac{1}{2}}\curl
a_{{a}} -\sigma_{{{a}}{{b}}}\omega^{{b}}=0 \;.
\end{equation}
Shear propagation:
\begin{equation}\label{1+3Shear}
\dot{\sigma}_{\langle {{a}}{{b}} \rangle }
+{\textstyle\frac{2}{3}}\Theta\sigma_{{{a}}{{b}}}
-\prjnab_{\langle {{a}}}a_{{{b}}\rangle }- a_{\langle
{{a}}}a_{{{b}}\rangle}
+\sigma_{{c}\langle {{a}}}\sigma_{{{b}}\rangle }{}^{c}+
\omega_{\langle {{a}}}\omega_{{{b}}\rangle} +E_{ab}-\frac{1}{2}\pi_{a b}
\,=0\;.
\end{equation}
Gravito-electric propagation:
\begin{eqnarray}\label{1+3GrEl}
\nonumber && \dot{E}_{\langle {{a}}{{b}} \rangle }+\frac{1}{2}\dot{\pi}_{\langle {{a}}{{b}} \rangle }
+\Theta  \left(E_{{{a}}{{b}}}-\frac{1}{6}{\pi_{a b}}\right) -\curl H_{{{a}}{{b}}}
+{\textstyle\frac{1}{2}}(\mu+p)\sigma_{{{a}}{{b}}}+\frac{1}{2}\prjnab_{\langle a}{q}_{b\rangle}+a_{\langle a}{q}_{b\rangle}\\&&
-2a^{c}\eta_{{c}{d}({{a}}}H_{{{b}})}{}^{d} -3\sigma_{c\langle a}
\left(E_{b\rangle }{}^{c}-\frac{1}{6}\pi_{b\rangle }{}^{c}\right) -\omega^{c}
\eta_{{c}{d}({{a}}}\left(E_{b)}{}^{d}-\frac{1}{6}{\pi_{b)}{}^{d}}\right)=0\;.
\end{eqnarray}
Gravito-magnetic propagation:
\begin{eqnarray}\label{1+3GrMag}
 &&\nonumber\dot{H}_{\langle
{{a}}{{b}} \rangle } +\Theta H_{{{a}}{{b}}} +\curl E_{{{a}}{{b}}}-\frac{1}{2}\curl \pi_{{{a}}{{b}}}-
3\sigma_{{c}\langle {{a}}}H_{{{b}}\rangle }{}^{c} -\frac{3}{2}\omega_{\langle a}q_{b\rangle}
-\omega^{c}\eta_{{c}{d}\langle a}H_{b\rangle}{}^{d}\\&&
+2a^{c}\eta_{{c}{d}\langle a}E_{b\rangle}{}^{d}+\frac{1}{2}\eta_{{c}{d}\langle a}\sigma_{b\rangle}{}^{c}q^{d}=0\;.
\end{eqnarray}
Vorticity constraint:
\begin{equation}\label{1+3VorConstr}
\prjnab^{{a}}\omega_{{a}} -a^{{a}}\omega_{{a}} =0\;.
\end{equation}
Shear constraint:
\begin{equation}\label{1+3ShearConstr}
\prjnab^{{b}}\sigma_{{{a}}{{b}}}+\curl\omega_{{a}}
-{\textstyle\frac{2}{3}}\prjnab_{{a}}\Theta +2[\omega,a]_{{a}}+ {q}_a=0\;.
\end{equation}
Gravito-magnetic constraint:
\begin{equation}\label{1+3GrMagConstr}
 \curl\sigma_{{{a}}{{b}}}-\prjnab_{\langle {{a}}}\omega_{{{b}}\rangle  }
 -H_{{{a}}{{b}}}-2a_{\langle {{a}}}
\omega_{{{b}}\rangle  }=0 \;.
\end{equation}
Gravito-electric divergence:
\begin{eqnarray}\label{1+3GrElConstr}
&& \prjnab^{{b}} \left(E_{{{a}}{{b}}}-\frac{1}{2}{\pi_{a b}}\right)-{\textstyle\frac{1}{3}}\prjnab_{{a}}\mu +\frac{1}{3}{\Theta}\bar{q}_a-\frac{1}{2}{\sigma}_{a b} {q}^b
-[\sigma,H]_{{a}} +\frac{3}{2}\eta^{abc}\omega_b q_c
+3H_{{{a}}{{b}}}\omega^{{b}}=0\;.
\end{eqnarray}
Gravito-magnetic divergence:
\begin{eqnarray}\label{1+3GrMagDiv}
 &&{\prjnab^{{b}} H_{{{a}}{{b}}}+(\mu+p)\omega_{{a}}+\frac{1}{2}\curl q_a +[\sigma,E]_{{a}}+\frac{1}{2}[\sigma,\pi]_{{a}}
 +3\omega^{{b}} \left(E_{{{a}}{{b}}}-\frac{1}{6}{\pi_{a b}}\right)=0\;.}
\end{eqnarray}
Conservation Equations (twice contracted Bianchi identities)
\begin{eqnarray}\label{Cons}
&&\dot{\mu} + \,\Theta\,(\mu+p)+2 a_c q^c+\sigma_{a}^{b}\pi^{a}_{b}=0\;,\label{eq:cons1}\\
&&q^{\langle a\rangle}+\prjnab_b\pi^{ab}+\prjnab^{a}{p} + (\mu+p)\,a^{a}+\frac{4}{3}\Theta q^a+\sigma^{a}_{b}q^{b}+a_{b}\pi^{ab}+\eta^{abc}\omega_b q_c=0  \,.
\end{eqnarray}

In the equations above the spatial curl of a vector and a tensor is
\begin{equation}
(\curl\,X)^{a} = \eta^{abc}\,\prjnab_{b}X_{c}\,,\qquad \qquad (\curl\,X)^{ab} = \eta^{cd\langle a}\,\prjnab_{c}X^{b\rangle}\!_{d}\,,
\end{equation}
respectively.

Finally, $\omega_{{a}}=\frac{1}{2}\eta_{a}{}^{{b}{c}}\omega_{bc}$ and the
covariant  commutators are
\[
[X,Y]_{{a}}=\eta_{{{a}}{c}{d}}X^cY^d\;,\qquad [W,Z]_{{a}} =\eta_{{{a}}{c}{d}}W^{c}{}_{e} Z^{{d}{e}}\,.
\]
%%%%%%%%%%%%%%%%%%%%%%%%%%%%%%%%%%%%%%%%%%%%%%
\subsubsection*{The 1+3 equations for the conformal observer.}
%%%%%%%%%%%%%%%%%%%%%%%%%%%%%%%%%%%%%%%%%%%%%%
Let us now see how these equation look like in the conformal frame.

\noindent Expansion propagation (generalized Raychaudhuri equation):
\begin{eqnarray}\label{1+3eqRayCF}
&&\nonumber \bar{\Theta}^\dag+{\textstyle\frac{1}{3}}\bar{\Theta}^2 -\prjconfnab{}^a
\bar{a}_{{b}}- \bar{a}_{{b}}\bar{a}^{{a}}+2\bar{\sigma}_{ab}
\bar{\sigma}^{ab} -2\bar{\omega}_{{a}}\bar{\omega}^{{a}}+{\textstyle\frac{1}{2}}(\bar{\mu} +
3\bar{p}) =\\&& -\frac{3}{2}\left(\frac{\Upsilon^{\dag}}{\Upsilon}\right)^2+\frac{\prjconfnab{}^a\Upsilon\prjconfnab_a\Upsilon}{\Upsilon^2}+ \frac{3}{2}\frac{\Upsilon^{\dag\dag}}{\Upsilon}+\frac{1}{2}\frac{\Upsilon^{\dag}}{\Upsilon}\bar{\Theta}-\frac{1}{2}\frac{\prjconfnab{}^2\Upsilon}{\Upsilon^2}- \frac{3}{2}\frac{a_b\prjconfnab{}^b\Upsilon}{\Upsilon}\;.
\end{eqnarray}
Vorticity propagation:
\begin{equation}\label{1+3VorCF}
\bar{\omega}^\dag_{\langle {{a}}\rangle }
+\frac{2}{3}\bar{\Theta}\bar{\omega}_{{a}} -\frac{1}{2}\overline{\curl
a}_{{a}} -\bar{\sigma}_{{{a}}{{b}}}\bar{\omega}^{{b}}=0 \;.
\end{equation}
Shear propagation:
\begin{eqnarray}\label{1+3ShearCF}
&& \nonumber \bar{\sigma}^\dag_{\langle {{a}}{{b}} \rangle }
+{\textstyle\frac{2}{3}}\bar{\Theta}\bar{\sigma}_{{{a}}{{b}}}
-\prjconfnab_{\langle {{a}}}\bar{a}_{{{b}}\rangle }- \bar{a}_{\langle
{{a}}}\bar{a}_{{{b}}\rangle}
+\bar{\sigma}_{{c}\langle {{a}}}\bar{\sigma}_{{{b}}\rangle }{}^{c}+
\bar{\omega}_{\langle {{a}}}\bar{\omega}_{{{b}}\rangle} +\bar{E}_{{{a}}{{b}}}-\frac{1}{2}\bar{\pi}_{a b}
\,\\&&=\frac{1}{2}\sigma_{ab}\frac{\Upsilon^{\dag}}{\Upsilon}-\frac{1}{2}\frac{\prjconfnab_{\langle {{a}}}\prjconfnab_{{{b}}\rangle }\Upsilon}{\Upsilon}+\frac{1}{4}\frac{\prjconfnab_{\langle {{a}}}\Upsilon \prjconfnab_{{{b}}\rangle }\Upsilon}{\Upsilon}\;.
\end{eqnarray}
Gravito-electric propagation:
\begin{eqnarray}\label{1+3GrElCF}
\nonumber && \bar{E}^\dag_{\langle {{a}}{{b}} \rangle }+\frac{1}{2}\bar{\pi}^\dag_{\langle {{a}}{{b}} \rangle }
+\bar{\Theta}  \left(\bar{E}_{{{a}}{{b}}}-\frac{1}{6}\bar{\pi}_{a b}\right) -\curl \bar{H}_{{{a}}{{b}}}
+{\textstyle\frac{1}{2}}(\bar{\mu}+\bar{p})\bar{\sigma}_{{{a}}{{b}}}+\frac{1}{2}\prjconfnab_{\langle a}\bar{q}_{b\rangle}+\bar{a}_{\langle a}\bar{q}_{b\rangle}\\&&\nonumber
-2\bar{a}^{c}\bar{\eta}_{{c}{d}({{a}}}\bar{H}_{{{b}})}{}^{d} -3\bar{\sigma}_{c\langle a}
\left(\bar{E}_{b\rangle }{}^{c}-\frac{1}{6}\bar{\pi}_{b\rangle }{}^{c}\right) -\bar{\omega}^{c}
\bar{\eta}_{{c}{d}({{a}}}\left(\bar{E}_{b)}{}^{d}-\frac{1}{6}\bar{\pi}_{b)}{}^{d}\right)=\\&&\left(\bar{E}_{{{a}}{{b}}}-\frac{1}{2}\bar{\pi}_{a b}\right)\frac{\Upsilon^{\dag}}{\Upsilon}-\frac{3}{2}\bar{\eta}^{cd}{}_{\langle a}H_{b\rangle d}\prjconfnab_{c}\Upsilon-\frac{3}{2}\frac{\bar{q}_{\langle a}\prjconfnab_{b \rangle }\Upsilon}{\Upsilon}\;.
\end{eqnarray}
Gravito-magnetic propagation:
\begin{eqnarray}\label{1+3GrMagCF}
 &&\nonumber\bar{H}^\dag_{\langle
{{a}}{{b}} \rangle } +\bar{\Theta} \bar{H}_{{{a}}{{b}}} +\overline{\curl E}_{{{a}}{{b}}}-\frac{1}{2}\overline{\curl \pi}_{{{a}}{{b}}}-3\bar{\sigma}_{{c}\langle {{a}}}\bar{H}_{{{b}}\rangle }{}^{c} -\bar{\omega}^{c}
\bar{\eta}_{{c}{d}\langle a}\bar{H}_{b\rangle}{}^{d}+2\bar{a}^{c}\bar{\eta}_{{c}{d}\langle a}\bar{E}_{b\rangle}{}^{d}\\&&-\frac{3}{2}\bar{\omega}_{\langle a}\bar{q}_{b\rangle}-\frac{1}{2}\bar{\eta}_{{c}{d}\langle a}\bar{\sigma}_{b\rangle}{}^{c}\bar{q}^{d}= \frac{1}{2} \bar{H}_{ab}\frac{\Upsilon^{\dag}}{\Upsilon}-\frac{1}{2}\bar{\eta}^{cd}{}_{\langle a}\left(\bar{E}_{b\rangle d}-\frac{3}{2}\bar{\pi}_{b\rangle d}\right)\prjconfnab_{c}\Upsilon
\;.
\end{eqnarray}
Vorticity constraint:
\begin{equation}\label{1+3VorConstrCF}
\prjconfnab{}^{{a}}\bar{\omega}_{{a}} -\bar{a}^{{a}}\bar{\omega}_{{a}} =0\;.
\end{equation}
Shear constraint:
\begin{eqnarray}\label{1+3ShearConstrCF}
&&\nn\prjconfnab{}^{{b}}\bar{\sigma}_{{{a}}{{b}}}+\curl\bar{\omega}_{{a}}
-{\textstyle\frac{2}{3}}\prjconfnab_{{a}}\bar{\Theta} +2[\bar{\omega},\bar{a}]_{{a}} +\bar{q}_a=\\&&\frac{\bar{\sigma}_{ba}\prjconfnab{}^{b}\Upsilon}{\Upsilon}+\frac{\bar{\omega}_{ba}\prjconfnab{}^{b}\Upsilon}{\Upsilon}+\frac{1}{2}\frac{\Upsilon^\dag}{\Upsilon}\frac{\prjconfnab_{b}\Upsilon}{\Upsilon}-\frac{1}{3}\bar{\Theta}\frac{\prjconfnab_{b}\Upsilon}{\Upsilon}+\frac{\prjconfnab_{b}\Upsilon^\dag}{\Upsilon}
\;.
\end{eqnarray}
Gravito-magnetic constraint:
\begin{equation}\label{1+3GrMagConstrCF}
 \overline{\curl\sigma}_{{{a}}{{b}}}-\prjconfnab_{\langle {{a}}}\bar{\omega}_{{{b}}\rangle  }
 -\bar{H}_{{{a}}{{b}}}-2\bar{a}_{\langle {{a}}}
\bar{\omega}_{{{b}}\rangle  }=0\;.
\end{equation}
Gravito-electric divergence:
\begin{eqnarray}\label{1+3GrElConstrCF}
&& \nonumber \prjconfnab{}^{{b}}\left( \bar{E}_{{{a}}{{b}}}-\frac{1}{2}\bar{\pi}_{a b}\right)
-{\textstyle\frac{1}{3}}\prjconfnab_{{a}}\bar{\mu}+\frac{1}{3}\bar{\Theta}\bar{q}_a-
\frac{1}{2}\bar{\sigma}_{a b} \bar{q}^b-3\bar{H}_{{{a}}{{b}}}\bar{\omega}^{{b}} -[\bar{\sigma},\bar{H}]_{{a}}+\frac{3}{2}\bar{\eta}^{abc}\bar{\omega}_b\bar{q}_c
\\&&=\frac{1}{2}\left( \bar{E}_{{{a}}{{b}}}-\frac{1}{2}\bar{\pi}_{a b}\right)\frac{\prjconfnab_{a}\Upsilon}{\Upsilon}-\frac{1}{2}\bar{\mu}
\frac{\prjconfnab_{a}\Upsilon}{\Upsilon}+\frac{3}{2}\bar{q}_a
\frac{\Upsilon^\dag}{\Upsilon}\;.
\end{eqnarray}
Gravito-magnetic divergence:
\begin{eqnarray}\label{1+3GrMagDivCF}
 &&\prjconfnab{}^{{b}} \bar{H}_{{{a}}{{b}}}
+(\bar{\mu}+\bar{p})\bar{\omega}_{{a}}+\frac{1}{2}\overline{\curl q}_a  +[\bar{\sigma},\bar{E}]_{{a}}+\frac{1}{2}[\bar{\sigma},\bar{\pi}]_{{a}}
 +3\bar{\omega}^{{b}} \left(\bar{E}_{{{a}}{{b}}}-\frac{1}{6}{\bar{\pi}_{a b}}\right)\nn \\ &&=\frac{1}{2}\bar{H}_{ab}\frac{\prjconfnab_{a}\Upsilon}{\Upsilon}-
 \frac{1}{2}[\bar{q},\frac{\prjconfnab\Upsilon}{\Upsilon}]_{{a}}\;.
\end{eqnarray}
Conservation Equations
\begin{eqnarray}\label{ConsCF}
&&\bar{\mu}^\dag + \,\bar{\Theta}\,(\bar{\mu}+\bar{p})+2 \bar{a}_c \bar{q}^c+\bar{\sigma}_{a}^{b}\bar{\pi}^{a}_{b}=\frac{1}{2}(\bar{\mu}+3\bar{p})
\frac{\Upsilon^\dag}{\Upsilon}+\frac{1}{\Upsilon}\bar{q}_c\prjconfnab{}^c\Upsilon\;,
\label{eq:cons1CF}\\
&&\nonumber\bar{q}_{\langle a\rangle}^\dag+\prjnab_b\bar{\pi}_{a}^{b}+\prjnab_{a}\bar{p} + (\bar{\mu}+\bar{p})\,\bar
{a}_{a}+\frac{4}{3}\bar{\Theta}\bar{q}_a+\bar{\sigma}_{a b}\bar{q}^{b}+\bar{a}^{b}\bar{\pi}_{ab}+\bar{\eta}_{abc}\bar{\omega}^b \bar{q}^c=\\&&=\frac{1}{2}(\bar{\mu}-\bar{p})\frac{\prjconfnab_a\Upsilon}{\Upsilon} +\bar{q}_a\frac{\Upsilon^\dag}{\Upsilon}+\frac{1}{\Upsilon}\bar{\pi}_{ab}
\prjconfnab{}^b\Upsilon \,.
\end{eqnarray}
%%%%%%%%%%%%%%%%%%%%%%%%%%%%%%%%%%%%%%%%%%%%%%%%%%%%%%%%
\subsubsection*{The 1+3 equations for $f(R)$-gravity in the Einstein frame.}
%%%%%%%%%%%%%%%%%%%%%%%%%%%%%%%%%%%%%%%%%%%%%%%%%%%%%%%%%
In the following we give, for completeness, the 1+3 equation for $f(R)$-gravity in the Einstein frame. The ones  in the Jordan frame can be found in \cite{SantePert,StructForm}.  The thermodynamic quantities for the scalar field $\phi$ are defined as
\begin{subequations}\label{ThermCFphi}
\begin{flalign}
&\mu^{\phi}\,=\frac{1}{2}(\phi^\dag)^2+\,\frac{1}{2}\prjconfnab\,{}^a\phi\prjconfnab_a \phi+W(\phi)\;,\\
&p^{\phi}\,=\frac{1}{2}(\phi^\dag)^2\,-\frac{1}{6}\prjconfnab\,{}^a\phi\prjconfnab_a\phi-W(\phi)\;,\\
&q_a= -\phi^\dag\prjconfnab_a \phi\;,\\
 &\bar{\pi}_{ab}=\prjconfnab_{\langle
a}\phi\,\prjconfnab_{b\rangle}\phi \;,
\end{flalign}
\end{subequations}
and we will use them in order to make the following equations more compact.

\noindent Expansion propagation (generalized Raychaudhuri equation):
\begin{eqnarray}\label{1+3eqRayCFphi}
&&\nonumber \bar{\Theta}^\dag+{\textstyle\frac{1}{3}}\bar{\Theta}^2 -\prjconfnab{}^a
\bar{a}_{{b}}- \bar{a}_{{b}}\bar{a}^{{a}}+2\bar{\sigma}_{ab}
\bar{\sigma}^{ab} -2\bar{\omega}_{{a}}\bar{\omega}^{{a}} =\\&&-{\textstyle\frac{1}{2}}(\bar{\mu}^m +
3\bar{p}^m)e^{\left(-\sqrt{2/3}\phi\right)}-(\phi^\dag)^2+W(\phi) \;.
\end{eqnarray}
Vorticity propagation:
\begin{equation}\label{1+3VorCFphi}
\bar{\omega}^\dag_{\langle {{a}}\rangle }
+\frac{2}{3}\bar{\Theta}\bar{\omega}_{{a}} -\frac{1}{2}\overline{\curl
a}_{{a}} -\bar{\sigma}_{{{a}}{{b}}}\bar{\omega}^{{b}}=0 \;.
\end{equation}
Shear propagation:
\begin{eqnarray}\label{1+3ShearCFphi}
&& \nonumber \bar{\sigma}^\dag_{\langle {{a}}{{b}} \rangle }
+{\textstyle\frac{2}{3}}\bar{\Theta}\bar{\sigma}_{{{a}}{{b}}}
-\prjconfnab_{\langle {{a}}}\bar{a}_{{{b}}\rangle }- \bar{a}_{\langle
{{a}}}\bar{a}_{{{b}}\rangle}
+\bar{\sigma}_{{c}\langle {{a}}}\bar{\sigma}_{{{b}}\rangle }{}^{c}+
\bar{\omega}_{\langle {{a}}}\bar{\omega}_{{{b}}\rangle} +\bar{E}_{{{a}}{{b}}}
\,\\&&=\frac{1}{2} \bar{\pi}{}^m_{ab} e^{\left(-\sqrt{2/3}\phi\right)} +\frac{1}{2}\prjconfnab_{\langle {{a}}}\phi \prjconfnab_{{{b}}\rangle }\phi\;.
\end{eqnarray}
Gravito-electric propagation:
\begin{eqnarray}\label{1+3GrElCFphi}
\nonumber && \bar{E}^\dag_{\langle {{a}}{{b}} \rangle }
+\bar{\Theta} \bar{E}_{{{a}}{{b}}} -\curl \bar{H}_{ab}
-2\bar{a}^{c}\bar{\eta}_{{c}{d}({{a}}}\bar{H}_{{{b}})}{}^{d} -3\bar{\sigma}_{c\langle a}
\bar{E}_{b\rangle }{}^{c} -\bar{\omega}^{c}
\bar{\eta}_{{c}{d}({{a}}}\bar{E}_{b)}{}^{d}=\\&&\nn-\frac{1}{2}(\bar{\pi}{}^m)^\dag_{\langle a b \rangle }e^{\left(-\sqrt{2/3}\phi\right)}+\frac{1}{\sqrt{6}}\phi^\dag\bar{\pi}{}^m_{ab}
e^{\left(-\sqrt{2/3}\phi\right)}-\left(\prjconfnab_{\langle a}\phi\right)^\dag\prjconfnab_{b \rangle }\phi+\frac{\bar{\Theta}}{6}\left(\bar{\pi}{}^m_{a b}e^{\left(-\sqrt{2/3}\phi\right)}+\prjconfnab_{\langle a}\phi\prjconfnab_{b \rangle }\phi\right)\\&&\nn -\frac{1}{2}(\bar{\mu}^m+\bar{p}^m)\bar{\sigma}_{ab}-\frac{1}{2}\left[(\phi^\dag)^2+
\frac{1}{3}\prjconfnab_{a}\phi\prjconfnab{}^{a}\phi\right]\bar{\sigma}_{a b}-\frac{1}{2}e^{\left(-\sqrt{2/3}\phi\right)}\prjconfnab_{\langle a}\bar{q}{}^m_{b\rangle}+\frac{1}{\sqrt{6}}e^{\left(-\sqrt{2/3}\phi\right)}
\prjconfnab_{\langle a}\phi\, \bar{q}{}^m_{b\rangle}\\&&\nn -\frac{1}{2}\prjconfnab_{a}\phi^\dag\prjconfnab{}^{a}\phi-\phi^\dag\prjconfnab_{a}
\prjconfnab{}^{a}\phi-\bar{a}_{\langle a}\bar{q}{}^m_{b\rangle}e^{\left(-\sqrt{2/3}\phi\right)}-\phi^\dag\bar{a}_{\langle a}\prjconfnab_{b\rangle}\phi-\frac{1}{2}\bar{\sigma}{}^c_{\langle a}\bar{\pi}_{b\rangle c }{}^{m}e^{\left(-\sqrt{2/3}\phi\right)}\\&&-\bar{\sigma}{}^c_{\langle a}\prjconfnab_{\langle c}\phi\prjconfnab_{b \rangle }\phi-\frac{1}{6}\bar{\sigma}_{ab}\prjconfnab_{a}\phi\prjconfnab{}^{a}\phi -\bar{\omega}^{c}
\bar{\eta}_{c(a}^{d}\frac{1}{6}\bar{\pi}_{b) d}{}^{m}e^{\left(-\sqrt{2/3}\phi\right)} -\bar{\omega}^{c}
\bar{\eta}_{c(a}^{d}\frac{1}{6}\prjconfnab_{\langle a}\phi\prjconfnab_{b \rangle }\;.
\end{eqnarray}
Gravito-magnetic propagation:
\begin{eqnarray}\label{1+3GrMagCFphi}
 &&\nonumber\bar{H}^\dag_{\langle
{{a}}{{b}} \rangle } +\bar{\Theta} \bar{H}_{{{a}}{{b}}} +\overline{\curl E}_{{{a}}{{b}}}-3\bar{\sigma}_{{c}\langle {{a}}}\bar{H}_{{{b}}\rangle }{}^{c} -\bar{\omega}^{c}
\bar{\eta}_{{c}{d}\langle a}\bar{H}_{b\rangle}{}^{d}+2\bar{a}^{c}\bar{\eta}_{{c}{d}\langle a}\bar{E}_{b\rangle}{}^{d}=\\
&&\nn\frac{1}{2}\overline{\curl \pi^m}_{{{a}}{{b}}}e^{\left(-\sqrt{2/3}\phi\right)}-\frac{1}{\sqrt{6}}\eta_{cd\langle a} \pi_{b\rangle}\!^{d}\,\prjnab^{c}\phi+\frac{1}{2}\overline{\curl \pi^\phi}_{{{a}}{{b}}}+\frac{3}{2}\bar{\omega}_{\langle a}\bar{q}^m_{b\rangle}e^{\left(-\sqrt{2/3}\phi\right)}\\
&&-\frac{3}{2}\bar{\omega}_{\langle a}\phi^{\dag}\prjconfnab_{\rangle b} \phi+\frac{1}{2}\bar{\eta}_{{c}{d}\langle a}\bar{\sigma}_{b\rangle}{}^{c}\bar{q}_m^{d}e^{\left(-\sqrt{2/3}\phi\right)}-
\frac{1}{2}\bar{\eta}_{{c}{d}\langle a}\bar{\sigma}_{b\rangle}{}^{c}\phi^{\dag}\prjconfnab{}^{d} \phi\;.
\end{eqnarray}
Vorticity constraint:
\begin{equation}\label{1+3VorConstrCFphi}
\prjconfnab{}^{{a}}\bar{\omega}_{{a}} -\bar{a}^{{a}}\bar{\omega}_{{a}} =0\;.
\end{equation}
Shear constraint:
\begin{equation}\label{1+3ShearConstrCFphi}
\prjconfnab{}^{{b}}\bar{\sigma}_{{{a}}{{b}}}+\curl\bar{\omega}_{{a}}
-{\textstyle\frac{2}{3}}\prjconfnab_{{a}}\bar{\Theta} +2[\bar{\omega},\bar{a}]_{{a}} =-\bar{q}_a^m e^{\left(-\sqrt{2/3}\phi\right)}-\phi^\dag\prjconfnab_{b}\phi
\;.
\end{equation}
Gravito-magnetic constraint:
\begin{equation}\label{1+3GrMagConstrCFphi}
 \overline{\curl\sigma}_{{{a}}{{b}}}-\prjconfnab_{\langle {{a}}}\bar{\omega}_{{{b}}\rangle  }
 -\bar{H}_{{{a}}{{b}}}-2\bar{a}_{\langle {{a}}}
\bar{\omega}_{{{b}}\rangle  }=0 \;. %\frac{1}{2\Upsilon}\eta^{mp\langle a}\sigma^{b\rangle}_{p}\prjconfnab_{m}\Upsilon \;.
\end{equation}
Gravito-electric divergence:
\begin{eqnarray}\label{1+3GrElConstrCFphi}
&& \nonumber \prjconfnab{}^{{b}} \bar{E}_{{{a}}{{b}}}-3\bar{H}_{{{a}}{{b}}}\bar{\omega}^{{b}} -[\bar{\sigma},\bar{H}]_{{a}}=\\
&&\nonumber \frac{1}{2}\prjconfnab{}^{{b}}\bar{\pi}^m_{a b}\,e^{\left(-\sqrt{2/3}\phi\right)}+\frac{1}{\sqrt{6}}\bar{\pi}^m_{a b}\prjconfnab{}^{b}\phi+ \frac{1}{2}\prjconfnab{}^{{b}}\bar{\pi}^\phi_{a b}+\frac{1}{3}\prjconfnab_{{a}}\bar{\mu}^m\,e^{\left(-\sqrt{2/3}\phi\right)}-\frac{1}{\sqrt{6}}\mu^m\,e^{\left(-\sqrt{2/3}\phi\right)} \prjconfnab{}^{b}\phi+\\
&&\nonumber\frac{1}{3}\prjconfnab_{{a}}\bar{\mu}^\phi-\frac{1}{3}\bar{\Theta}\bar{q}^m_a\,e^{\left(-\sqrt{2/3}\phi\right)}-\frac{1}{3}\bar{\Theta}\bar{q}^m_a\,e^{\left(-\sqrt{2/3}\phi\right)}-\frac{3}{2}\bar{\eta}^{abc}\bar{\omega}_b\bar{q}^m_c\,e^{\left(-\sqrt{2/3}\phi\right)}+\frac{3}{2}\bar{\eta}^{abc}\bar{\omega}_b\phi^\dag\prjconfnab_c \phi\\&&
-\frac{1}{2}\bar{\sigma}_{a b} \bar{q}_m^b\,e^{\left(-\sqrt{2/3}\phi\right)}+\frac{1}{2}\bar{\sigma}_{a b}\phi^\dag\prjconfnab{}^b \phi\;.
\end{eqnarray}
Gravito-magnetic divergence:
\begin{eqnarray}\label{1+3GrMagDivCFphi}
 \nn&&\prjconfnab{}^{{b}} \bar{H}_{{{a}}{{b}}}+[\bar{\sigma},\bar{E}]_{{a}}
 +3\bar{\omega}^{{b}}\bar{E}_{{{a}}{{b}}}=\\&&\nn-(\bar{\mu}^m+\bar{p}^m)\bar{\omega}_{{a}}\,e^{\left(-\sqrt{2/3}\phi\right)}-(\bar{\mu}^\phi+\bar{p}^\phi)\bar{\omega}_{{a}}-\frac{1}{2}\overline{\curl q^m}_a\,e^{\left(-\sqrt{2/3}\phi\right)}+\frac{1}{\sqrt{6}}\eta_{a}^{b c}\bar{q}^m_b\prjconfnab_c\phi-\frac{1}{2}\overline{\curl q^\phi}_a\\&&-\frac{1}{2}[\bar{\sigma},\bar{\pi}^m]_{{a}}\,e^{\left(-\sqrt{2/3}\phi\right)}- \frac{1}{2}[\bar{\sigma},\bar{\pi}^\phi]_{{a}}-\frac{1}{2}\bar{\omega}^{{b}} {\bar{\pi}^m_{a b}}\,e^{\left(-\sqrt{2/3}\phi\right)}-\frac{1}{2}\bar{\omega}^{{b}} {\bar{\pi}^\phi_{a b}}\;.
\end{eqnarray}
Conservation Equations for standard matter (as a general fluid):
\begin{eqnarray}\label{ConsCFphi}
&&\bar{\mu}_m^\dag +\prjconfnab{}^c\bar{q}^m_c+ \,\bar{\Theta}\,(\bar{\mu}^m+\bar{p}^m)+2 \bar{a}_c \bar{q}_m^c+\bar{\sigma}_{a}^{b}\bar{\pi}^{a}_{b}=\frac{1}{\sqrt{6}}(\bar{\mu}+3\bar{p})\phi^\dag+2\sqrt{\frac{2}{3}}\bar{q}_c^m\prjconfnab{}^c\phi\;,\label{eq:cons1CFphi}\\
&&\nonumber\bar{q}_{\langle a\rangle}^\dag+\prjnab_b\bar{\pi}_{a}^{b}+\prjnab_{a}\bar{p} + (\bar{\mu}+\bar{p})\,\bar
{a}_{a}+\frac{4}{3}\bar{\Theta}\bar{q}_a+\bar{\sigma}_{a b}\bar{q}^{b}+\bar{a}^{b}\bar{\pi}_{ab}+\bar{\eta}_{abc}\bar{\omega}^b \bar{q}^c=\\&&=\frac{1}{\sqrt{6}}(\bar{\mu}-\bar{p})\prjconfnab_a\phi+\sqrt{\frac{2}{3}}\phi^\dag\bar{q}_a+\sqrt{\frac{2}{3}}\bar{\pi}_{ab}\prjconfnab{}^b\phi \,.
\end{eqnarray}

\section*{Acknowledgments}
SC wishes to thank Prof.~P.K.S.~Dunsby, Dr.~J.~Larena and Dr.~A.~Hamilton  for useful discussion. SC was funded by Generalitat de Catalunya through the Beatriu de Pinos contract 2007BP-B1 00136. The work has been also supported in part by MEC (Spain), projects FIS2006-02842 and PIE2007-50/023, and by AGAUR, contract 2005SGR-00790.

\end{document}